\documentstyle[aaspp4,flushrt]{article}

\def\et{{\it et \thinspace al.}\ }
\begin{document}
\title{
THE STAR FORMATION HISTORY OF THE STARBURST REGION NGC 2363 AND 
ITS SURROUNDINGS
\altaffilmark{1,2}
}

\author{LAURENT DRISSEN, JEAN-REN\'E ROY, 
CARMELLE ROBERT and DANIEL DEVOST\altaffilmark{3}}

\affil{D\'epartement de Physique, Universit\'e Laval and Observatoire du mont
M\'egantic\\
Qu\'ebec, QC, G1K 7P4, Canada\\
Electronic mail: ldrissen, jrroy, carobert, ddevost@phy.ulaval.ca
}

\author{REN\'E DOYON}

\affil{D\'epartement de Physique, Universit\'e de Montréal
and Observatoire du mont M\'egantic\\
Montr\'eal, QC, H3C 3J7, Canada\\
doyon@astro.umontreal.ca
}

\altaffiltext{1}{Based on observations with the NASA/ESA
Hubble Space Telescope, obtained at the Space Telescope Science
Institute, which is operated by AURA, Inc., under NASA contract NAS5-26555}

\altaffiltext{2}{Based on observations obtained at the
Canada-France-Hawaii Telescope, which is
operated by the National Research Council of Canada, the Centre National
de la Recherche Scientifique de France, and the University of Hawaii}

\altaffiltext{3}{Also at Space Telescope Science Institute, 3700 San Martin 
drive, Baltimore, MD, 21218}

\begin{abstract}
We present {\it Hubble Space Telescope}  optical images and UV spectra, as well as
ground-based near-infrared images of the
high surface brightness giant H\,{\sc ii} region NGC 2363 (NGC 2366-I)
and its surroundings.
The massive star content of the southern end of the dwarf irregular
galaxy NGC 2366 is investigated, with an emphasis on Wolf-Rayet
and red supergiant stars, and we attempt the reconstruction of the
time sequence of the most recent episode of massive star formation
at the southwestern tip of the galaxy. The ages of the clusters are
respectively 10 Myr for NGC 2366-II, 2.5-5 Myr for NGC 2363-B and less than
1 Myr for NGC 2363-A. In particular,
we show that the most massive super cluster
A of NGC 2363 is still embedded in dust; from the photoevaporative erosion
or ``cleaning'' time scale of the associated cloud, we infer its age to be 
$\sim 10^6$ yr or less. We conclude that the
star-forming complex NGC 2366-I and II is a good example of a multiple stage
starburst with a characteristic age decreasing from 10 Myr to less
than 1 Myr over a linear scale of 400 pc. 
The age sequence of the stars and the gas kinematics suggest that 
these powerful star formation episodes are being triggered by a small 
passing-by satellite. 
\end{abstract}

\keywords{ISM: individual (NGC 2363) --- galaxies: individual (NGC 2366) ---
stars: emission line, Be --- stars: evolution}

\section{INTRODUCTION}

One of the most powerful local starburst events is taking place in
the nearby magellanic galaxy NGC 2366. This activity is betrayed by three
giant H\,{\sc ii} regions, one of them having its own NGC number
(NGC 2363 = Mkn 71) and being the highest surface brightness H\,{\sc ii} region in the sky.
By its stellar content and star formation rate, NGC 2363 competes with
30 Doradus in the Large Magellanic Cloud (Kennicutt 1984). Therefore
NGC 2363 has been the target of multiple studies, and many of
its properties in the ultraviolet and optical domains are well known; its oxygen
abundance is about 1/10 solar (Peimbert, Pe\~na, \& Torres-Peimbert 1986;
González-Delgado \et 1994; although see Luridiana 1998 for a different
interpretation of the data leading to a higher abundance), where 12 + log O/H$_\odot$ = 8.87 (Grevesse, Noels, \& Sauval 1996). It 
has a well determined carbon abundance (Garnett \et 1995).
Its shows unusual kinematic features:
an expanding supershell (Roy \et 1991) and hypersonic gas
(Roy \et 1992; González-Delgado \et 1994). Only a few Wolf-Rayet
stars are present; they are associated with
the eastern cluster of NGC 2363 that we will also call knot B 
(Drissen, Roy, \& Moffat 1993, hereafter DRM). Surprisingly, as we will 
show, we do not
see the stars of the main ionizing cluster, called knot A, in NGC 2363.
A rare Luminous Blue Variable, located
5$''$ east of this cluster, has been undergoing a 
spectacular eruption since about 1994 (Drissen, Roy, \& Robert 1997); this star,
NGC 2363-V1,
is presently the brightest optical source in the galaxy. We
are monitoring it spectroscopically with the Hubble Space
Telescope (HST program 7391).
The rich and complex system of ionized gas filaments and shells in
the southwestern of NGC 2366 has been investigated
by Hunter \& Gallagher (1997) and Martin (1998). Martin has derived ages for the largest  supershells ($\tau$ = 0.6 R/$v$) as 7.9 Myr for the supershell east of NGC 2366-I and
II and 5.7 Myr for the shell north of NGC 2366-I (= NGC 2363). Hunter \& 
Gallagher have infered older ages.

In this paper, we analyse optical images and ultraviolet spectra obtained with 
the Hubble Space Telescope (HST), as well as near infrared CFHT images, to 
understand the intense star formation  in the three main
H\,{\sc ii} regions of NGC 2366, identified as NGC 2366 I (= NGC 2363), II and III
(Figures 1 and 2). In particular, we will demonstrate that the southwestern
tip of NGC 2366 is undergoing a passing wave of star formation, moving
from east to west, and that the most recent burst is so
young that its stellar supercluster is still dust embedded;
its stars are not yet seen in the ultraviolet or optical domains. We will 
explore the starbursting process by dating 
the propagating star formation process in the galaxy, and will discuss
the possibility of a trigger by a small interacting satellite.

\section{OBSERVATIONS}

\subsection{Imagery}

Most of the images of NGC 2366 used in this work were obtained in 1996 January
with the WFPC2 camera aboard the {\it Hubble Space Telescope}; the giant H\,{\sc ii}
region NGC 2363 is located at the center of the high resolution PC1 CCD.
At the adopted distance of 3.44 Mpc (Tolstoy \et 1995), the
scale is $\sim$17 pc/arcsec (i.e. 0.8 pc/pixel on the PC1).
The broadband F439W ($\sim$ Johnson B; 3 exposures of 900 s) and 
F547M ($\sim$ Stromgren y; 2 $\times$ 800 s) filters were used. The F547M 
filter was prefered to the wider, more standard F555W filter, because it 
excludes the strong [O\,{\sc iii}] $\lambda\lambda$4959-5007 nebular lines. However, because of
the high surface brightness of the NGC 2363 nebula, dust scattered stellar light
and diffuse
nebular {\it continuum} contaminates both F439W and F547M images.
Narrowband images of the same regions were obtained with the following 
filters: F656N (H$\alpha$; 2 $\times$ 500 s), to study the distribution of 
the ionized gas and F469N (He\,{\sc ii} $\lambda$4686; 5 $\times$ 2000 s) to 
detect Wolf-Rayet stars and a possible diffuse nebular component.
During our monitoring of NGC 2363-V1, we obtained two series of short
exposures covering NGC 2363 with filters F170W, F336W, F547M and F108M
in 1998 March and December. Only the brightest stars are seen in these
images, but they were used to determine the flux of clusters A and B.
We have used DAOPHOT (Stetson 1987) to derive instrumental magnitudes
and colors of the stars, and the prescription of Whitmore (1995) and
Holtzmann \et (1994) to transform these into the standard photometric system.

In addition to the HST observations, a 10 minute $H\alpha$ image of
NGC 2366 was secured with the Loral $2K \times 2K$ CCD of the
Canada-France-Hawaii telescope (CFHT) in 1996 February.  The $H\alpha$
image, shown in Figure 1, outlines the location of the WFPC2 field.
The WFPC2 frame, a composite of the F547M, F439W and F656N images, is
displayed in Figure 2.  We also obtained JHK images with the MONICA
infrared camera (Nadeau \et 1994) attached at the f/8 focus of the
CFHT in 1997 January. The K band image is displayed in the bottom part
of Figure 3 with the high resolution PC1 image (top part). Basic data
reduction was performed for the infrared images using software developed by
the Montr\'eal group (D. Nadeau). The images were flux calibrated using
the IR UKIRT standard stars FS2, FS21, and FS25.
The uncertainties in the JHK magnitudes are $\sim 0.15$ mag. For a detailed
description of the infrared reduction steps and calibration, see Devost (1999).

\subsection{Spectroscopy}

HST {\it Faint Object Spectrograph} spectra of NGC 2363's knots A and B
were gathered in 1996 December, with the G130H grating
(covering the 1140 to 1606 \AA\ range with a resolution of 1 \AA\ per diode) 
and the 0.86$''$-wide circular aperture. 
The observations were obtained as follows. The January 1996 WFPC2 images
were used to identify a bright (V=17) isolated star $80''$ 
north of NGC 2363; its position and offsets from the 
two targets were then measured with the IRAF/STSDAS task METRIC.
This star was centered in the FOS aperture with a series of ACQ/BIN and ACQ/PEAK
acquisition sequences, leading to a pointing accuracy of about 0.04 arcsecond
(Keyes \et 1995). The telescope was then slewed to cluster A and the 
observation took place. The telescope was slewed back to the offset star, 
re-centered in the aperture before coming back to cluster B. 
The total integration times were 6150 s for cluster 
A and 9730 s for cluster B. 

Moreover, we obtained a series of HST/STIS long slit
spectrograms of NGC 2363-V1 in
1997 November with the G140L and G230L gratings in order to study the
physical parameters of this erupting LBV. The spectra of V1 will be
discussed elsewhere (Drissen \et 1999a), but two bright stellar
objects (stars or unresolved stellar aggregates) within NGC 2363-B
fell by chance into the slit and provide us with valuable information about
the stellar content of this cluster (section 4.3).

\section{THE STARBURST COMPLEX NGC 2366-I AND II: AN OVERVIEW}

The massive star population of NGC 2366, with the notable exception
of the core of NGC 2363, is generally well resolved into individual
stars in our images (Figure 2). The F547M image reaches V$\sim$ 25.5;
this corresponds to $M_v \sim -2.5$, or spectral type B2 ($M \sim 10 M_\odot$)
on the main sequence.
On a deep H$\alpha$ image (Figure 1), NGC 2366-I and II appear like a
single, high surface brightness ionized gas complex. But the high
resolution continuum images reveal that they are two distinct entities
with their own star-formation history.
NGC 2366-III, discussed in more detail in section 5, appears like
a group of loose OB associations without a well-defined core. 
The WFPC2 (F547M) and MONICA (K-band) images of NGC 2366-I and II
are displayed in Figure 3. Red supergiant stars identified in the
color-magnitude diagrams (with $B - V \geq 1.0$ and
$M_V \leq -5.0$ in Figure 4) are circled. We also note the presence
of a slightly resolved source, at the eastern edge of NGC 2363, which is bright 
in the K image, but barely detected in the J and H images. The near-IR colors
of this object are those of hot ($\sim 500$ K) dust.

The spatial distribution of the red supergiants is reflected in the
two color-magnitude diagrams: the difference between NGC 2363 and
NGC 2366-II is striking. Only three red supergiants
are seen in the CMD of NGC 2363; in fact, these stars are
located outside the main body of the H\,{\sc ii} region and may not be physically
related to it. 
In contrast, more than two dozen red supergiants 
are seen within the boundaries of NGC 2366-II. The surface 
density of these stars is much higher in NGC 2363-II than in the rest of the
galaxy, attesting to their association with the H\,{\sc ii} region. Their 
red supergiant 
nature is also confirmed by the MONICA J, H and K images and the
near-infrared two-color diagram (Figure 5); this diagram will be discussed
in more details below.
The presence of such a large number of red supergiant stars in a giant
H\,{\sc ii} region is uncommon and suggests
that the main burst of star formation in NGC 2366-II
occured some 7 to 10 Myrs ago
(Mayya 1997); this is consistent with the age of supershell A (Martin 1998).
On the other hand, many bright, blue stars are seen in the western part
of NGC 2366-II, close to the bright H$\alpha$ ridge visible in Figure 2.
These are certainly younger than most red supergiants seen in the
eastern section of the cluster; their presence suggests that the 
star formation process in NGC 2366-II did not occur in a single,
instantaneous burst. We are probably seeing here a slightly more evolved
version of the two-stage starburst NGC 2363 itself (see next section).

\section{NGC 2363}

In ground-based continuum images, NGC 2363 presents a double morphology 
with two intensity maxima that we identified (DRM) as western knot (A) 
and eastern knot (B). The new HST images now resolve B into an elongated,
bright (V$\sim$18), slightly resolved core surrounded by many dozens blue
stars (Figure 6). From now on we shall refer to the OB association,
within the boundaries of the expanding cavity, as ``cluster B'', and the
compact object at its core as ``knot B''.
The western half of NGC 2363 contains few visible stars and has a very
high H$\alpha$ surface brightness. What appears at its core, knot A,
remains unresolved into stars; we 
will show below that because of dust we do not see stellar light directly. 
Actually, the different nature of knots A and B is also made clear in 
the near-infrared two-color diagram (Figure 5):
we see that knot B has near-IR colors of hot, blue stars, while the colors
of knot A are very similar to those of a pure H\,{\sc ii} region (see
Figure 5 in Doyon \et 1995).

\subsection{The Population of Wolf-Rayet Stars}

Wolf-Rayet stars in cluster B of NGC 2363 were first detected
by DRM via narrow-band filter imagery.
The excess of light at $\lambda$4686 was then attributed to the presence
of about 5 WR stars, although a contribution from a {\it nebular} component
could not be excluded and was indeed suggested by the images. 
Long-slit spectroscopy
(Gonz\'alez-Delgado \et 1994) confirmed both the WR and nebular signatures:
apart from a strong, {\it narrow} He\,{\sc ii} $\lambda$ 4686 emission line of nebular
origin, broad bumps at 4660 \AA\ and 5800 \AA, typical of WR stars,
are detected. The ratio of the 4650 and 5800 \AA\ bumps indicates that
an early-type WC star dominates the WR population in cluster B.

The WFPC2 images provide a clearer picture.
In order to identify the WR stars, we subtracted the continuum
image (a weighted average of the F547M and F439W images)
from the F469N image (see Drissen \et 1999b for more details); the result 
is shown in Figure 7. The quality of the image subtraction can be judged 
by the very low residuals at the position of knot A. The largest 
{\it absolute} residuals on the net image are at the position of V1, which 
is by far the 
brightest star in the field; they amount to $\pm$ 5\% of the peak value of 
the stellar PSF in the continuum image. The total net counts within the PSF 
of V1 in the net image is, within 1 $\sigma$, identical to the total counts
in a nearby region of the sky.

Three stars in cluster B clearly have an excess of 
light at 4686 \AA ; a fourth one is detected 8.6$''$ (140 pc) north 
of the cluster. WR1 has a continuum magnitude (M$_v \sim -5.5$) typical of 
Galactic WR stars. The $\lambda$ 4686 excess detected at the northwestern
tip of the elongated core of cluster B (WR2) could be the result of an 
imperfect image subtraction due to the complex distribution of light
in this region, but all our attempts to make it disappear by shifting
the images around before subtraction failed, and we are confident that it
is real and betrays the presence of a WR star. The star with the strongest 
$\lambda$ 4686 excess, WR3, is very likely the WC star which dominates the 
global WR spectrum of the region.
With $M_v \sim -7$, this star appears much brighter than any known WC star,
but it is very likely that we are in fact observing an
unresolved stellar aggregate which includes one WC star.
Finally, the isolated WR4 has a weak, but significant excess, suggesting that
it might be a narrow-lined WNL star. We note that a
weak residual also appears in the net 4686 image of DRM (see 
their Figure 3) at the same position as WR4, but was not discussed in their
paper. WR4 is located 150 pc (projected distance) from the center of cluster
B and its association with the later is not clear.

No WR candidate, nor significant diffuse He\,{\sc ii} $\lambda$4686 emission are
detected near cluster A, nor in NGC 2366-II.
It is worth emphasizing that no other WR candidate has been detected
anywhere else in the galaxy despite a very careful search, using both the
subtraction technique and by blinking the continuum and ``on-line'' F469N
frames. Using the same techniques and similar exposure times, 
we detected a large number of
WR candidates in NGC 2403 (about 40 per WFPC2 field), a spiral galaxy 
at the same distance as NGC 2366 (see Drissen \et 1999b, where incompleteness 
is also discussed in detail).

The small number of WR stars in and around NGC 2363 is very surprising
at first sight. Giant H\,{\sc ii} regions with similar H$\alpha$ luminosities
such as 30 Dorarus in the LMC or NGC 2403-I and II, host between one and
two dozen WR stars each. NGC 604 and NGC 595 in M33 also have a substantial
WR population (Drissen, Moffat \& Shara 1993).
This paucity of WR stars in NGC 2363 
can be explained by two factors. The metallicity
of NGC 2363 has generally been determined to be close to one tenth solar
(Peimbert \et 1986). At this low metallicity, very few O stars (those more massive than
$\sim$ 80 M$_\odot$) are expected to shed enough mass via their wind
to reach the WR phase (Schaerer \& Vacca 1998). But Luridiana (1998) 
suggests that the metallicity of this region has been underestimated, and that a
value of $Z = 0.25 Z_\odot$ would give a better fit to the observed
line ratios. The higher value is also consistent with the
mean abundance level established
by Roy \et (1996) for the other regions of NGC 2366.
If this value is adopted, then the number of WR stars
ought to be comparable to that of 30 Dor. Nevertheless, we think that the
age spread within NGC 2363 provides a better explanation: as we will
explain below, cluster A provides the bulk of the ionizing flux for the
nebula, but we do not see a signature of WR stars because (a) it is
much too young to harbor a sizeable WR population and (b) its stars
are still hidden in dust.

\subsection{The He\,{\sc ii} Nebular Excess and its origin}

Ground-based narrow-band imagery suggest the
presence of a spatially extended nebular He\,{\sc ii}$\lambda$ 4686 emission
($l \sim $ 4\arcsec) in NGC 2363; this has been confirmed spectroscopically
(see references above). In order to check if this diffuse
component was also detected in the WFPC2 images, we first subtracted the
point sources (i.e. WR candidates) from the net $\lambda$ 4686 image shown
in Figure 7, using DAOPHOT's substar task. 
The resulting image was convolved with a gaussian profile
of 1.5 pixel radius to increase the signal-to-noise ratio. 
Figure 8 presents the monochromatic H$\alpha$
image of NGC 2363, with the contours of nebular He\,{\sc ii}$\lambda$ 4686
emission superposed. The total flux from this diffuse component is
$F_{neb}^{4686} = 2.5 \times 10^{-15} erg~cm^2~s^{-1}$, a value which is
compatible with the spectroscopic data of González-Delgado \et (1994).
The faint extended zone of nebular He\,{\sc ii}$\lambda$4686
emission corresponds to an obvious cavity
in overall emission and its orientation is the same as that of the
``chimney'' discovered by Roy \et (1991). Garnett \et (1991) have
reviewed the possible 
ionization mechanisms for nebular He\,{\sc ii} emission. They are:

\noindent 1. {\it Hot stellar ionizing continua.--} Classical
nebulae ionized by O stars having T$_{eff} \leq$ 55,000K do not
produce strong He$^{++}$. However models for
massive stars  with mass loss (Maeder \& Meynet 1987) predict
evolution blueward to T$_{eff} \geq$ 70,000 K. Pakkull (1991) suggest that very
early WN-type stars do have T$_{eff}$ high enough to excite He\,{\sc ii}
emission. WO stars could have very high temperatures and be capable
of producing the He$^{++}$ zone, but the spectrum of cluster B (González-Delgado
\et 1994) does not show the broad OVI $\lambda$3811,34 lines
 characteristic of
WO stars. Schaerer \& Vacca
(1998) show that nebular He\,{\sc ii} emission should be associated
with WC/WO stars and hot WN stars evolving to become
WC/WO stars. WO stars would show very strong and broad P Cygni profiles
in the O\,{\sc vi} lines. WNE would also show high excitation features
such as P\,{\sc v} and S\,{\sc vi} but no O\,{\sc vi} lines.

\noindent 2. {\it Shock excitation.--} Shocks can produce
relatively strong He\,{\sc ii} emission when shock velocities are
$\sim$ 120 km/s or higher (Binette, Dopita \& Tuohy 1985; Sutherland, Bicknell, \&
Dopita 1993). However ground-based spectra 
do not reveal evidence for shocks, and the mean velocity of the
expanding supershell revealed by Roy \et (1991) is about 45 km/s.

\noindent 3. {\it X-rays. --} X-ray sources could have enough flux
extending into the extreme UV to produce He$^{++}$ nebulae
(Pakull \& Angebault 1986). No X-ray source is known to exist
in NGC 2363; but it could always be hypothesized that the He$^{++}$
zone is a fossil from an X-ray source which switched off within
the past century (Garnett \et 1991).

We have shown that 3 WR stars are present at the core of
cluster B. None of them is a WO, but the global spectrum being
one of a hot WC4 star, we suspect that they supply a good fraction
of the hard UV-flux
which is the source of the nebular He\,{\sc ii} emission in NGC 2363. 
Observations with FUSE could help to identify these stars. 

\subsection{Selected Stars}

NGC 2363-V1 is, since early 1995, the most luminous star in its galaxy.
Once too faint to be detected in ground-based images, it is now
$M_v \sim -10.5$. Analysis of its light curve and spectrum shows that
it is a Luminous Blue Variable experiencing a major eruption,
probably similar to that of $\eta$ Carina in the middle of the 19$^{th}$
century (Drissen \et 1999a). The luminosity of V1 prior to its current
eruption was much lower than that of $\eta$ Car, so it is likely
to be older (age $\sim$ 4 to 5 Myrs)
and less massive than its Galactic counterpart.

By chance, the long slit we used to obtain the STIS spectrum of V1
intercepted the light from two stellar objects (identified as stars
1 and 2 in Figures 5 and 6). While the image of star 1 appears unresolved,
star 2 is obviously multiple. Both objects are too bright to be
single stars, and their spectra are likely to be composite. These
are shown in
Figure 9, along with those of stars having similar spectral morphology
in the Small Magellanic Cloud. Actually, these SMC templates are
averages of a few stars having the corresponding spectral type,
a part of the spectral library built by C. Robert.
Evolutionary synthesis models predict an age of 3 to 5 Myr for a cluster 
with an O5~II star spectrum and 4.5 to 7 Myr for a cluster with an
O9~III star spectrum.

We also note the presence of two blue stars surrounded by H\,{\sc ii} shells,
located about 100 pc south of knot A (see Figure 6). The shells have
a diameter (2 and 6 pc, respectively) similar to those of Galactic LBVs.
These objects warrant further spectroscopic investigation.

\subsection{The FOS Ultraviolet Spectra of Knots A and B}

The FOS UV spectra of knots A and B are presented in Figure~10.
To increase the signal-to-noise ratio, the data were smoothed using
a boxcar filter over four pixels (the original spectral resolution is
maintained as the spectra were oversampled by quarter-diode subpixel steps).
The FWHM of simple interstellar lines (S{\sc ii}\,1254,
S{\sc ii}+Si{\sc ii}\,1260, O{\sc i}+Si{\sc ii}\,1303, and Si{\sc ii}\,1527)
and of the Ly$\alpha$ geocoronal line reveal an average spectral resolution 
$\lesssim 2$~\AA, consistent with the expected resolution of the G130H
grating.
In order to properly interpret the spectral synthesis, the spectra have 
been shifted to position
the interstellar lines at their rest frame wavelength. Shifts of
-190 and -150~km~s$^{-1}$ were applied to the spectra of knots A and B,
respectively. Given the known systemic velocity of NGC 2363, +110 ($\pm 10$)
~km~s$^{-1}$ (Izotov, Thuan \& Lipovetsky 1997), the remaining velocity (i.e.
80 and 40~km~s$^{-1}$ for A and B, respectively) is mainly due to
the uncertainty in measuring the zero-point of the instrument wavelength 
scale, which is typically of the order of 130~km~s$^{-1}$.

The spectrum of knot B reveals strong P~Cygni profiles at 1550~\AA\ 
(the C{\sc iv} doublet at 1548.2 and 1550.8~\AA) and at
1240~\AA\ (the N{\sc v} doublet at 1238.8 and 1242.8~\AA). These are typical
signatures of fast and dense stellar winds from massive stars
(see the UV atlas of O stars by Walborn, Nichols-Bohlin \& Panek 1985). 
Many UV resonance lines, like C{\sc iv} 
and N{\sc v}, are also formed in stellar photospheres
and the interstellar medium, in which case they 
appear as absorption lines only without an emission component. 
A narrow interstellar contribution to C{\sc iv} 
is indeed seen superposed on the wind profile of knot B. 

Another important clue to the stellar content of knot B is the fact that the
Si{\sc iv}\,1400 line (at 1393.8 and 1402.8~\AA), also a resonance doublet,
does not show a wind P~Cygni profile in this spectrogram. 
Because the Si$^{+3}$ ion has a smaller opacity than C$^{+3}$,
Si{\sc iv}, unlike C{\sc iv}, develops a P~Cygni profile only
in the denser winds of giant and supergiant stars, and remains a photospheric
absorption line in main sequence hot stars.
The weak absorption line at 1371~\AA\ (OV; see Walborn \et 1985)
is typical of mid-temperature main 
sequence to giant O stars; this line  is seen as a broad P~Cygni line in 
O3 stars or O supergiants and disappears in late O and B stars.
If late B stars were the main contributors to the UV lines of knot B, their
photospheric absorption lines, which are broader than their counterparts
in O stars or the interstellar medium, would show up in the spectrum; this
is not the case.

This first qualitative look at the spectrum of knot B already tells us that 
this stellar population must be dominated by massive, young O stars 
at a time before most of them have evolved into giants or supergiants.
Knot B must therefore be young (i.e. a few Myr only), as we will determine
more quantitatively from the UV line synthesis below.

The spectrum of knot A is strikingly different from that of knot B: it is
much flatter and does not harbor P Cygni wind profiles.
There is no sign of stellar wind signatures at C{\sc iv}\,1550,
N{\sc v}\,1240 nor Si{\sc iv}\,1400, nor indications of cooler B stars 
with broad photospheric lines.
The C{\sc iv}\,1550 feature in knot A is actually a strong narrow nebular 
emission, consistent with the very high nebular surface density seen in
the H$\alpha$ image. 
An HST/FOS spectrum of knot A covering the wavelength range from 
1600 to 2300~\AA\ (Garnett \et 1995) also shows many nebular 
emission lines of O{\sc iii}]\,1666, Si{\sc iii}]\,1883, 1892, and 
C{\sc iii}]\,1909. The slope of the continuum over the whole region
covered by the FOS is consistent with a purely nebular continuum and stellar light
scattered by dust.
We do not associate the weak continuum depression blueward of the C{\sc iv}
emission with a wind signature, but rather attribute it to
a sequence of many absorption lines (of Si{\sc ii} and possibly
Fe{\sc iv}) of interstellar origin.
We therefore come to the surprising, but interesting, conclusion that 
{\it the massive young stars responsible for the photoionization of 
knot A and its surrounding are not detected in the ultraviolet
spectrum and are therefore completely hidden from view at this 
wavelength}.
 
\subsection{Spectral Synthesis of the UV spectrum of knot B}

We perform the UV spectral synthesis of knot B using the code of
Dionne (1999). This code is a revised version of the code of
Leitherer, Robert \& Drissen (1992) and a slightly different version of the Starburst99 code
(Leitherer \et 1999). The Dionne code is optimized for UV spectral
synthesis at various metallicities (and for the treatment of massive close
binaries). It contains a library of UV spectra (normalized to a continuum equal
to unity) at medium-high resolution of O, B and WR stars in the solar
neighbourhood and the lower metallicity environment of the Small and Large
Magellanic Clouds. It uses the recent evolutionary tracks
of the Geneva group (Charbonnel \et 1993, Schaller \et 1992, Schaerer \et 1993a, b,
Meynet \et 1993), interpolated to the metallicity of the 
Magellanic Clouds. 
Stellar atmosphere models (Kurucz 1992 for OB stars, Schmutz, Leitherer
\& Gruenwald 1992 for WR stars)
are used to flux-calibrate the UV library 
spectra. The code follows a population of stars where at each time step
the stellar evolutionary tracks give, among other paramaters, 
the effective temperature 
and luminosity of each star which are used to assign a spectral type, and 
then a UV spectrum from the library.
  
At first, we synthesize the UV spectra of knot B for a stellar population
at low metallicity with the evolutionary tracks at $0.1~Z_\odot$
and the SMC UV library (built with HST UV spectra of O stars from the SMC
and IUE spectra of WR and B stars in the solar environnement).
A Salpeter type IMF is considered, i.e. a power law with an
exponent $\alpha = 2.35$, with the lower mass limit 
$M_{low} = 1~M_\odot$ and upper mass limits $M_{upp} = 30$ to $120~M_\odot$.
The star formation rate is either constant, i.e. new stars are created at
each time step, or instantaneous, i.e. all stars are formed at an initial time.
 
For the instantaneous models, the best agreement between
the synthesized wind lines
and the observation is found for $M_{upp} = 60$ to $120~M_\odot$
and the corresponding ages $\tau = 3$ to 2~Myr.
With a higher upper mass limit, younger models are favored because
the absorption part of the wind profiles of Si{\sc iv} and C{\sc iv}
becomes too deep too fast.
If the upper mass limits is bellow $60~M_\odot$, the models do not predict
enough wind signatures in C{\sc iv}.
For older ages, the wind contribution predicted for Si{\sc iv} is too strong.
These wind effects disappear with the death of the massive stars.
Around 7~Myr, Si{\sc iv} becomes a broad absorption, 
C{\sc iv} is a weak P~Cygni profile (formed in B supergiants), 
N{\sc V} is absent and photospheric lines of Si{\sc iii} around 1300~\AA\ 
are predicted. 
N{\sc v} is not very well reproduced with the best models (i.e. the
predicted strength of the P~Cygni emssion is not large enough at young ages),
but this line has a lower weight due to its strong blending with Ly$\alpha$
which is not well reproduced by the library spectra.
 
In the case of the continuous models, the age is slightly stretched by 0.5-1~Myr 
and the upper mass limit lower value increased to $80~M_\odot$.
Indeed, since the best fits are found for a young age, the new stars added 
in the continuous case are not numerous enough yet to dilute the wind 
signatures. Other models have been considered with various IMF parameters. 
Good fits are found for a lower IMF exponent or an increased lower mass  value
(i.e. when the massive stars are more numerous relative to the low mass
stars). But this has little effect on the age and the upper mass cutoff.
If the metallicity is increased  (e.g. $Z = 0.25~Z_\odot$ with 
the LMC UV library), the strength of the wind profiles are strong at an 
earlier age and for a lower upper mass limit, but then it becomes impossible 
to reproduce simultaneouly the Si{\sc iv} and C{\sc iv} lines.
 
Figure~11 shows a superposition of the observed spectrum of knot B
with synthetic spectra at different ages for an instantaneous model 
with $\alpha = 2.35$ and $M_{upp} = 80~M_\odot$. The observed spectrum is 
dereddened so the UV continuum slope reproduces the theoretical 
slope for a young burst at low metallicity (i.e. with the UV flux
$F \propto \lambda^\beta$, $\beta = -2.7$; Leitherer \et 1999).
Using the SMC extinction law of Prévot \et (1984), a galactic contribution
$E(B-V)_{Gal} = 0.04$, we find the  small internal extinction 
$E(B-V)_{Internal} = 0.018$. 
The best fit to the data, corresponding to an age of 2.5 Myr, is 
shown superposed on the observed spectrum in Figure~11. As discussed already, 
younger models do not show the strong wind line of C{\sc iv}. Older
models give, at first, too much wind signatures in Si{\sc iv}. 
With the death of the most massive stars, the wind profiles disappear 
and broad B stars absorptions appear.
As we can see in Figure~11, none of the best fits are perfect in the center
of C{\sc iv}. The residual between the models and the observation shows
a large emission component of unknown origin
with a velocity width (FWZI) of $\sim 2000$~km~s$^{-1}$,
centered on C{\sc iv}.
 
Assuming a distance of 3.44~Mpc, the synthesis code allows to estimate,
based on the best line models, that about 800 B stars and 40 O stars are
present in knot B. 
Three WR stars are seen from images in narrow bands. 
However, no WR stars are predicted by the population synthesis models that
best match the UV spectra: the first
WR star to appear at a metallicity of $0.1~Z_\odot$, when stars
as massive as $120~M_\odot$ are present, is at around 2.8~Myr.
On the one hand, the evolutionary tracks for the massive stars we
use are still in developement. For example, rotation effects 
are being included which will possibly modify the time at which the
first WR star will appear (Maeder 1999). Furthermore, as we are studying 
a small population, fluctuation in the IMF must be expected.

We now compare the light and energy output predicted by the population
synthesis model (age=2.5 Myr, Salpeter IMF, Z = 0.1 $Z_\odot$) that
best reproduces the UV spectra of knot B with the observations.

\begin{itemize}

\item{\it V magnitude -}
The predicted unreddened flux at 5500~\AA\ 
is $3 \times 10^{-16}$~erg~s$^{-1}$~\AA$^{-1}$~cm$^{-2}$. From the WFPC2
images, we measure a total magnitude $ V = 18.5$ within a radius of
0.43$''$ (to match the FOS aperture) centered on knot B. This corresponds
to $2.2 \times 10^{-16}$~erg~s$^{-1}$~\AA$^{-1}$~cm$^{-2}$ if we adopt
$A_v = 0.4$, in relatively good agreement with the population synthesis
models. 

\item{\it Ionizing photons -}
The total number of Lyman continuum photons produced by
massive stars in knot B is $Log N_{Lyc} = 50.78$, that is about 6\%
of the number required to ionize the entire nebula. This suggests that
the most important source of ionization in NGC 2363 comes from
knot A.

\item{\it Kinetic energy -}
Finally, the total energy returned to the ISM since the starburst
2.5 Myrs ago via stellar winds is of the order of $5 \times 10^{50}$ erg
(we note here that this energy is a factor of 10 smaller than the same
stars would generate in a solar metallicity environment; see Leitherer
\et 1999). 
This is 2 orders of magnitude smaller than the kinetic energy in the
expanding bubble discovered by Roy \et (1991, 1992). Since the bubble
cannot have been blown by the winds of massive stars in the core of cluster B,
we conclude that the energy released by the winds of massive stars and 
especially supernova explosions within the bubble but outside the 
knot B are responsible for the bubble expansion. As we showed
in section 4.3, there are hints that the stellar population surrounding
knot B is a bit older than the core itself, an age segregation
that we also observe in the super star clusters of NGC 2403 (Drissen \et 1999b).

\end{itemize}

\subsection{Dust Embedded Super Stellar Cluster of Massive Stars at Knot A}

All the information we have gathered so far, namely the existence of 3 WR
stars, the spectral features of stars 1 and 2 and the core, the absence
of red supergiants, and the kinematical age of the bubble, indicate that
cluster B is 2.5 to 5 Myr old, with a real age spread between the core
and the periphery. We now show that cluster A is even younger.

The western half of NGC 2363, surrounding knot A, has a high H$\alpha$
luminosity and exceptional surface brightness. It must therefore be ionized
by a rich young cluster of massive stars characterized by a UV continuum sharply
rising towards shorter wavelengths and full of P Cygni profiles. This is
obviously not what we see: the absence of stellar features in the 
flat UV spectrum of NGC 2363-A strongly suggest that the super star cluster
responsible for most of the ionization of NGC 2363 is
completely shrouded in dust. Despite the low apparent
extinction derived from the Balmer decrement (c(H$\beta) \sim$0.2),
the presence of dust in the nebula is obvious even in the optical domain. 
NGC 2363 is a very high surface brightness object and its
electron temperature at T = 15 000 K is unusually high (Kennicutt, Balick,
\& Heckman 1980; Peimbert \et 1986; González-Delgado
\et 1994). Thus we can expect the nebular continuum to be
relatively strong, especially in the ultraviolet where the
contribution of the two-photon continuum becomes important.
Dust particles scatter continuous
radiation of the stars immersed in the H\,{\sc ii} region, resulting
in an observable diffuse continuum (Osterbrock 1989).
The main contributions to the nebular continuum in the F547M band are
the hydrogen recombination continuum, the He+ recombination continuum
(about 1/10 of the hydrogen one) and the 2-photon continuum;
the fraction of He++ is only 0.003 of H+ (González-Delgado et al. 1994).
We make
the assumption that the ratio of these continua to H$\alpha$ is about the
same in the H$\alpha$ and F547M filter bandpasses. This is justified
by the fact that the sum the H, He recombination continua
and of the two-photon at 15~000 K
are roughly constant between the two wavelengths (Aller 1984).
The nebular continuum can then be roughly approximated
by scaling the H$\alpha$ image. The F547M filter should be free of all major
nebular line emission.
Therefore, we used this image to subtract a scaled H$\alpha$ image
obtained with the F656N filter (after
removing the stellar profiles in both the F547M and F656N HST images
using the standard techniques of the {\sc daophot} package
in IRAF). The remaining diffuse emission, in particular in and around knot A,
is most likely due to stellar light scattered by dust; the result is
shown in Figure 12. If on the larger scale, the nebular continuum probably
dominates, in the brightest regions of knot A the dust scattered continuum
may be up to four times stronger than the nebular continuum.

Let us now use another important piece of information:
NGC 2363 shows as a point source in the
IRAS Point Source Catalog with S$_{12\mu m}$ $<$ 0.25 Jy, S$_{25\mu m}$ = 0.7235 Jy,
S$_{60\mu m}$ = 3.303 and S$_{100\mu m}$ = 4.578 Jy. The flux value at 12.5$\mu$m is
an upper limit.  Although the IRAS spatial resolution is crude, we will show
that knot A is probably the point source.

Dust embedded massive stars have
remarkably similar far-infrared (FIR) flux
density distributions independent of distances (Wood and Churchwell 1989a,b).
On color diagrams Log [S$_{100\mu m}$/S$_{60\mu m}$] vs
Log [S$_{25\mu m}$/S$_{12\mu m}$] and Log [S$_{60\mu m}$/S$_{12\mu m}$] vs
Log [S$_{25\mu m}$/S$_{12\mu m}$], the FIR colors of NGC 2363 are very close to
the locus of FIR colors of
ultracompact H\,{\sc ii} regions (Wood \& Churchwell 1989a). Since the
measurement at 12$\mu$m is an upper limit, then the colors
are pushed well into the regime of ultracompact H\,{\sc ii} regions.
Observations of dust in galactic compact H\,{\sc ii} regions
imply visual extinctions between 25 and 1250 mag (Chini \et 1986 a,b).

Figure 13 presents the spectral energy distribution (SED) of the core of
cluster B and of knot A, from the FOS spectra, WFPC2 and MONICA images
(all within an aperture of 0.86$''$ to match the FOS aperture), and
the IRAS data. We assume here that most of the FIR flux of NGC 2363 
originates from knot A, which is a reasonable assumption given the
arguments presented above.
The SED of cluster B is very well fitted by an instantaneous starburst
model of age 3 Myr, but the SED of knot A is very different. The 
flat UV to optical distribution and the huge FIR excess are in fact
reminiscent of the SED of dust-embedded starburst galaxies (Silva \et 1998),
although the starburst in NGC 2363 is much younger that those in
the galaxy sample of these authors.

We have therefore strong evidence that the region is shrouded by dust: super 
cluster A is in
the ultracompact H\,{\sc ii} region stage, that is that the newly formed 
stars are still embedded in their natal molecular clouds (Churchwell 1991). 
The far-infrared flux can then be used to infer the star formation
rate and the dust content of the region.
Following Devereux \& Young (1990a)
\begin{equation}
L (40 - 120 \mu{\rm m}) = 3.65 \times 10^5 \ (2.58 S_{60\mu{\rm m}} + S_{100\mu{\rm m}})\ D^2 \ [L_\odot],
\end{equation}
where $S_{60\mu{\rm m}}$ and $S_{100\mu{\rm m}}$ are the IRAS 60 and
100 $\mu$m flux densities in units of Jy, and D is the distance of the object
in Mpc. For NGC 2363, L(40 $-$ 120 $\mu$m)
= 5.7 $\times 10^7$ L$_\odot$. The star formation rates for the high-mass end  are
infered from the FIR luminosity (Devereux \& Young 1991;
Sauvage \& Thuan 1992), as
\begin{equation}
SFR_{\rm FIR} (\geq 10 M_\odot) = 1.4 \times 10^{-10} L_{\rm FIR} (L_\odot)
\ [{\rm M_\odot \ yr^{-1}}],
\end{equation}
or from the H$\alpha$ luminosity (Kennicutt 1983), as
\begin{equation}
SFR_{H\alpha} (\geq 10 M_\odot) = 5.45 \times 10^{-9} L_{H\alpha} (L_\odot)
\ [{\rm M_\odot\  yr^{-1}}].
\end{equation}
With the appropriate values for NGC 2363, $SFR_{\rm FIR}$ $\simeq$ 0.01 M$_\odot$
yr$^{-1}$, and $SFR_{H\alpha}$ = 0.02 M$_\odot$ yr$^{-1}$
 for high mass stars. Both rates are comparable, in agreement with
the hypothesis that the high-mass stars which ionize the
hydrogen gas generate the far-infrared luminosity (Devereux \& 
Young 1990a).

The total mass of dust in NGC 2363 can also be estimated from
the far-infrared luminosity, although several effects and uncertainties
may affect the result (see Draine 1990).
If we assume that the NGC 2363 far infrared luminosity is dominated by a
single grain component and temperature, we have for dust
heated by OB stars
following Devereux \& Young (1990b)
\begin{equation}
M_d = 4 a \rho S_{100\mu{\rm m}} D^2/3 Q_{100\mu{\rm m}} B_{100\mu{\rm m}}(T),
\end{equation}
where $B_{100\mu{\rm m}}(T)$ is the value of the Planck function
at $100\mu$m, $S_{100\mu{\rm m}}$ is the 100 $\mu$m flux density,
$D$ is the distance, $Q_{100\mu{\rm m}}$ is the grain emissivity
at $100\mu$m, $a$ is the grain radius, and $\rho$ is the grain
density. Using appropriate values for the grain
parameters (Hildebrand 1977, Hildebrand \et 1983), the above equation is simplified to
\begin{equation}
M_d = C S_{100\mu{\rm m}} D^2 (e^{144/T} - 1) M_\odot,
\end{equation}
where $D$ is the distance in Mpc, $S_{100\mu{\rm m}}$ is the 100 $\mu$m flux density in Jy and $T$ is the dust temperature. $C$ is a poorly
known constant
which depends on the grain opacity which can vary between about 1.8 and 4.58
(Eales, Wynn-Williams, \& Duncan 1989).
The total mass of dust derived is obviously very sensitive to the
adopted temperature. Adopting
$T_d \sim 30$ K (Churchwell, Wolfire, \& Wood 1990),
$C \sim 4$ and $D = 3.44$ Mpc, one finds
$M_d \sim 2.6 \times 10^4\ M_\odot$ of dust. If we assume
$M_{H_2} = 100\ M_d$ (Solomon \et 1997), the mass of molecular
hydrogen associated with NGC 2363 is of the order of $10^6\ M_\odot$.
Although this figure appears reasonable, one must be aware
of the associated uncertainties; this naive approach is likely
to underestimate the total dust mass (Draine 1990).

One can use this estimate of dust-gas mass and
the fact that the stars of super cluster A are not directly
visible to infer some upper limit to its age by
deriving its ``cleaning time scale'' (Blaauw 1991).
We will suppose that most of the obscuring  part is
in the dense accretion disks out of which the individual stars formed.
We will assume that the dominant extinction is due
to the individual dusty disks, and that the gas and dust
are being eroded by the steady replenishment of the explanding
plasma by material photoevaporating from neutral disks orbiting
the massive stars. Using the integrated observed properties of NGC 2363 
(Kennicutt 1984), we take  the number of ionizing photons from
the young massive stars as 10$^{52}$ s$^{-1}$ corresponding to
a total mass of  50~000 M$_\odot$ in OB stars; the ionized mass
of hydrogen  is $\sim 10^6$ M$_\odot$. Disk photoevaporation
around single young stars has been studied and modeled by
Hollenbach \et (1994). Using and scaling their
global relations, which do not
depend on the details of the accretion disk-star geometry,
we can get a rough estimate of the lifetime of
the dust and gas cloud wrapping super cluster A. The photoevaporative mass-loss rate in the
strong wind case, which is the most appropriate case for NGC 2363-A, is
given by
\begin{equation}
\dot{M} \simeq 6 \times 10^{-5} \ \Phi_{49}^{0.44}\ v_{w8} \ \ [M_\odot {\rm yr}^{-1}],
\end{equation}
\noindent where $\Phi_{49} = \Phi_i/10^{49}$ s$^{-1}$ is the photon
rate, and $v_{w8}$ is
the stellar wind speed in units of 1000 km s$^{-1}$. 
The total effective
photoevaporative rate in the surroundings of the massive stars
of NGC 2363 would be of the order of $\geq 10^{-3}$ M$_\odot$ yr$^{-1}$. Thus
10$^3$-10$^4$ M$_\odot$ could be ablated in 1 Myr,
a value consistent with the 1\% fraction of
the molecular gas residing in dense
cloud cores out of which the circumstellar disks form. In
a super-cluster, each star probably has its own individual accretion disk.
Hollenbach et al. estimate
that lifetimes of $\geq 10^5$ yr are probably achieved by 2-10 M$\odot$
disks, a range of disk masses corresponding
to the critical value of $\sim 0.3$ M$_*$ above wich the disks are
unstable. The effect of dust will lengthen these lifetimes,
but in a cluster of stars-disks, the boiling away of the
remnant accretion disks will be accelerated by the first
stars emerge from their cocoons. Taking these various
effects into consideration,
a cleaning time scale of about 1 Myr is plausible.
We surmise that super stellar cluster A cannot be much older
than this.
There are likely additional 
larger dust envelopes or shells somewhat decoupled from the individual stellar disks
(Churchwell \et 1990).
These larger cocoon with lower densities than disks 
may rapidly develop a clumpiness which allows easy paths for photons to get through,
as seen in young galactic H\,{\sc ii} regions (Cox, Deharveng, \& Leene
1990).

\section{NGC 2366-III}

The WFPC2 image of the central part of NGC 2366-III is shown in Figure 14,
and the color-magnitude diagram of the same region is displayed in Figure 15.
Fifteen red supergiant stars are located within the boundaries of
the H\,{\sc ii} region;
their surface density there is a factor of 5 higher than in the remaining of the
WF CCD, attesting to their association with NGC 2366-III. It is
noticeable that most of the RSG stars are located at the periphery of
the central region, where the density of ionized gas (including a
possible wind-blown bubble) is higher.

\section{GAS MORPHOLOGY AND KINEMATICS}

Ground-based H$\alpha$ imagery of NGC 2366 betrays a spectacular network
of filaments and shells (Hunter \& Gallagher 1997; Martin 1998).
 Martin (1998) has used longslit echelle
spectroscopy to study these structures; she calculated various
parameters (age and total kinetic energy) for the most obvious filament structures
surrounding NGC 2363; she derived kinematic ages less than 10 Myr. 
A slightly different view is possible. Basing their arguments on models of supershell 
formation, Hunter \& Gallagher (1997) believe that the massive
stars that were the cause of the filaments cannot be
those at the core of the H\,{\sc ii} regions; they suggest that these gaseous
structures must be from a previous episode of star formation  as old as 50 Myr.
Ground-based [O\,{\sc iii}]$\lambda$5007 Fabry-Perot interferometry done
at CFHT (some of the data were presented in Roy \et 1991) is helpful
in revealing the large scale kinematics of the ionized gas. For example,
apart from the unusual properties of the gas in NGC 2363, the [O\,{\sc iii}]$\lambda$5007
line profiles are very narrow and symmetrical in NGC 2366-II and III. While the
mean heliocentric velocity of the gas is about 90-100 km/s in the
``quiescent'' region of NGC 2363 (we exclude the supershell
and hypersonic velocity gas), and 100-110 km/s in NGC 2366-II, it is 60-70 km/s
in NGC 2363-III. There is a systematic gradient between NGC 2366-I and II
to III in the diffuse [O\,{\sc iii}] emission, with heliocentric velocities
 decreasing from 100 to 80 km/s.

The H\,{\sc i} maps of Braun (1995) display a similar velocity
range. Morever the H\,{\sc i} column density map shows a rim of gas
of lower velocity gas west
of the main optical body of the galaxy, and NGC 2366-III is clearly
associated with this rim. The velocity contours reveal a disturbed shape
at the southwestern tip
and show the same velocity difference of 30-45 km/s between NGC
2366-III and the two other large H\,{\sc ii} regions. These kinematical features
suggest that NGC 2366-III may belong to a system which is kinematically 
disconnected from the main body of NGC 2366. We will surmise
that it may be a small satellite cloud associated with NGC 2366,
whose capture or interaction is responsible for triggering star formation in
that region of the galaxy.

\section{STARBURST TRIGGERING BY INTERACTION: A DISCUSSION}

The observations presented above clearly point to a well-determined
age sequence running from east to west among the star clusters. The typical 
age of the stellar population range from 10 Myr (NGC 2366-II) to
3-5 Myr (NGC 2363-B) to less than 1 Myr (NGC 2363-A). Although these events
could be completely uncorrelated, it seems interesting to examine
the possibility that a single event led to the formation of these clusters.

The relative locations and velocities of NGC 2366-III and of
the associated rim of H\,{\sc i} gas support the scenario
of a satellite cloud being now in close interaction
with the main body of the galaxy NGC 2366. 
NGC 2366-III is at a distance of 1.4 kpc, in the plane of 
the sky, of the mid-point between
NGC 2363 and NGC 2366-II. Supposing a space velocity of 150 km/s,
a realistic value for a satellite of a magellanic galaxy,
NGC 2366-III was very close to the southern
tip of the galaxy less than 10 Myr ago, and could have triggered the 
interaction which has led to the strong episodes of star
formation in the southern half of NGC 2366 and in NGC 2366-III. 
This environment of a starbursting magellanic galaxy is not unique. Actively star forming 
dwarf galaxies often (in more than 50\% of the
cases searched) have nearby companions (Taylor 1995; 
Taylor \et 1995). For example, Stil \& Israel
(1998) have recently found a low-mass H\,{\sc i} companion to the
post-starburst galaxy NGC 1569 at a projected distance of
5 kpc. The velocity difference between NGC 1569 and
NGC 1569-H\,{\sc i} is 40 km/s. 
Furthermore the age sequence of the super-star clusters is
consistent with an interaction scenario as a trigger to the
starburst episodes. 

It would be illuminating to know at what phase of the interaction
the strongest starburst activity is expected to occur. Barnes \&
Hernquist (1996) state that the timing of an interaction-induced
starburst depends on the relative orientations of the two
disks and on the pericentric separation at first close approach.
Stronger tidal responses in disks are associated with closer
passages (or direct collisions) than wider or retrograde
encounters. Also the response to a tidal perturbation
depends on the internal structure of the galaxy.

\section{CONCLUSIONS}

We have presented images and spectra of the southern end of the bar in
NGC 2366, covering the wavelength range from 115 nm to 2.2 $\mu$m. From
these data, we deduce the following:

1. The star-forming complex NGC 2366-I and II is a clear example of a 
multiple-stage starburst, with characteristic ages decreasing from
$\sim$ 10 Myr (NGC 2366-II) to $\sim$ 3 - 5 Myr (NGC 2363-B) to $\leq$
1 Myr (NGC 2363-A).

2. The large number of red supergiants in NGC 2366-II indicates that the
bulk of stars formed $\sim$ 10 Myr ago, but the presence of a few bright
blue stars close to an intense ridge of ionized hydrogen suggests that
star formation went on until more recently ($\sim$ 5 Myr ago).

3. NGC 2363-B contains no red supergiants, but numerous massive O stars,
three WR candidates and an erupting LBV, and is located in the middle
of an expanding superbubble of dynamical age 3 Myr. 
The core of cluster B is composed of young (2.5 to 3 Myr) massive stars
surrounded by a slightly older (4-5 Myr) population.

4. The kinetic energy released by the winds of massive stars in the
core of cluster B (within the FOS aperture) is not sufficient
(by a factor of $\sim 100$) to have blown the expanding superbubble.

5. NGC 2363-A is a very young ($\leq$ 1 Myr) and dense star cluster
still embedded in dust. A significant fraction of the UV radiation
emitted by its hidden massive stars must leak out of the clumpy
dust cloud to provide the bulk of the ionizing flux of the giant 
H\,{\sc ii} region.

6. NGC 2366-III is composed of multiple loose OB associations. We
suggest that it may be part of a satellite of NGC 2366, whose passage
close to the end of the bar some 10 Myr ago initiated the multiple-stage
starburst in NGC 2366-II and NGC 2363.

If it were at a distance of 100 Mpc of more, the whole star-forming
complex would be less than 1 arcsecond in diameter. The interpretation
of its global parameters in terms of
multiple starburst episodes would then be hampered by the lack
of spatial resolution. 
Finally, our study of the small starburst NGC 2363
demonstrates vividly how powerful a tool HST has been
at helping to unveil the stellar content and evolution
of neaby giant star forming regions. The stunning 
progress in our understanding of giant H\,{\sc ii} regions
can be fully measured by reading the pionneering work
on NGC 2363 by Kennicutt \et (1980).

\vskip 2.0truecm
We thank Fran\c cois St-Pierre who did the analysis and derived
the parameters of the distant Sc I spiral galaxy seen through the
disk of NGC 2366 and Mario Leli\`evre who helped with Figure 12.
This investigation was funded in part by the Natural Sciences
and Engineering Research Council of Canada, by the Fonds FCAR of
the Government of Qu\'ebec and by Universit\'e Laval.
{\bf Most figures in the astro-ph archives are compressed and dithered.
The original figures are available at the following URL:
http://astrosun.phy.ulaval.ca/astro/N2363.html}

\section{Appendix: Distant Background Galaxies}

Table 1 gives the list and coordinates of the background galaxies 
detected in our WFPC2 V images of the field of NGC 2366, in increasing
order of right ascension. Images of these galaxies will be posted on
the NED database. The most striking object is the
spiral N2366BG7, which has a magnitude V $\sim$ 17.1 and B $-$ V $\approx$ 1.0
(Figure 16). We used this galaxy to derive the reddening through the 
disk at that location in NGC 2366,
and we have estimated its distance. We assumed an intrinsic value for the
color of the seen-through spiral first by comparing it to nearby morphological counterparts. We adopted B $-$ V = 0.51, by taking the average color
of NGC 628, NGC 5660, NGC 6070 and NGC 6118, fair representatives of
Sc I galaxies. Assuming a Galactic extinction of A$_{\rm V}$ = 0.13
along this line of sight (Burstein \& Heiles 1984), and an internal
extinction due to the inclination of the galaxy of 0.16,
the derived extinction due to NGC 2366 is  A$_{\rm V}$(N2366) $\approx$ 0.35.
The K-correction (0.28) was done assuming a redshift of 0.13 based on
a distance of $\sim$500 Mpc. The distance was inferred again from a comparison with
the apparent
diameters of the four sibling galaxies.

\newpage
\begin{deluxetable}{lcccl}
\tablenum{1}
\tablecaption{Background Galaxies in the field of NGC 2366}

\tablehead{
\colhead{Object} &
\colhead{$\alpha$ (2000)} &
\colhead{$\delta$ (2000)} &
\colhead{V} &
\colhead{Comment} \\
}
\startdata

N2366BG1&07 28 38.7&69 12 19.2&23.4& \nl
N2366BG2&07 28 39.0&69 11 47.3&23.4& \nl
N2366BG3&07 28 43.0&69 12 02.2&21.2& \nl
N2366BG4&07 28 43.2&69 12 39.3&21.9& \nl
N2366BG5&07 28 43.2&69 12 46.1&23.0& \nl
N2366BG6&07 28 44.2&69 12 05.0&23.1& \nl
N2366BG7&07 28 45.3&69 12 19.2&17.1& Scd (see Figure 16)\nl
N2366BG8&07 28 45.9&69 12 40.9&22.6& \nl
N2366BG9&07 28 47.9&69 12 40.0&21.4& \nl
N2366BG10&07 28 51.1&69 11 21.1&21.6& \nl
N2366BG11&07 28 51.3&69 12 02.9&21.3& \nl
N2366BG12&07 28 53.3&69 11 21.0&21.9& \nl
N2366BG13&07 28 54.5&69 11 12.0&19.3& Interacting with N2366BG14?\nl
N2366BG14&07 28 54.6&69 11 11.3&21.1& Interacting with N2366BG13?\nl

\enddata
\end{deluxetable}

\newpage

\begin{figure}
\figurenum{1}
\plotone{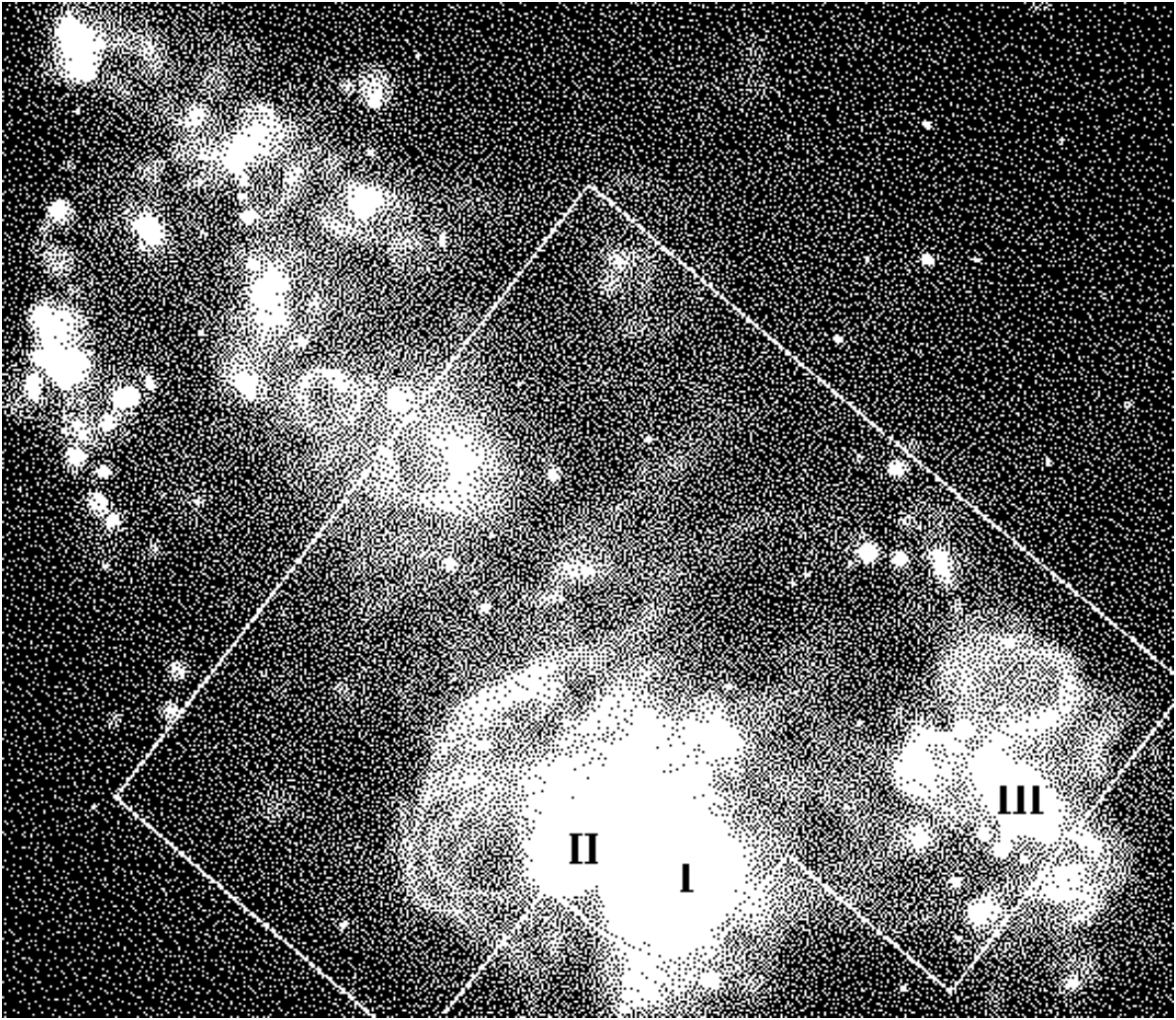}
\figcaption{H$\alpha$ image of NGC 2366 from the CFHT. The HST/WFPC2 field
of view is outlined, and the giant H\,{\sc ii} regions are numbered. The field
is $230'' \times 195''$ (3.8 kpc $\times$ 3.2 kpc at the distance of NGC 2366).
North is up, east to the left.
}
\end{figure}

\begin{figure}
\figurenum{2}
\plotone{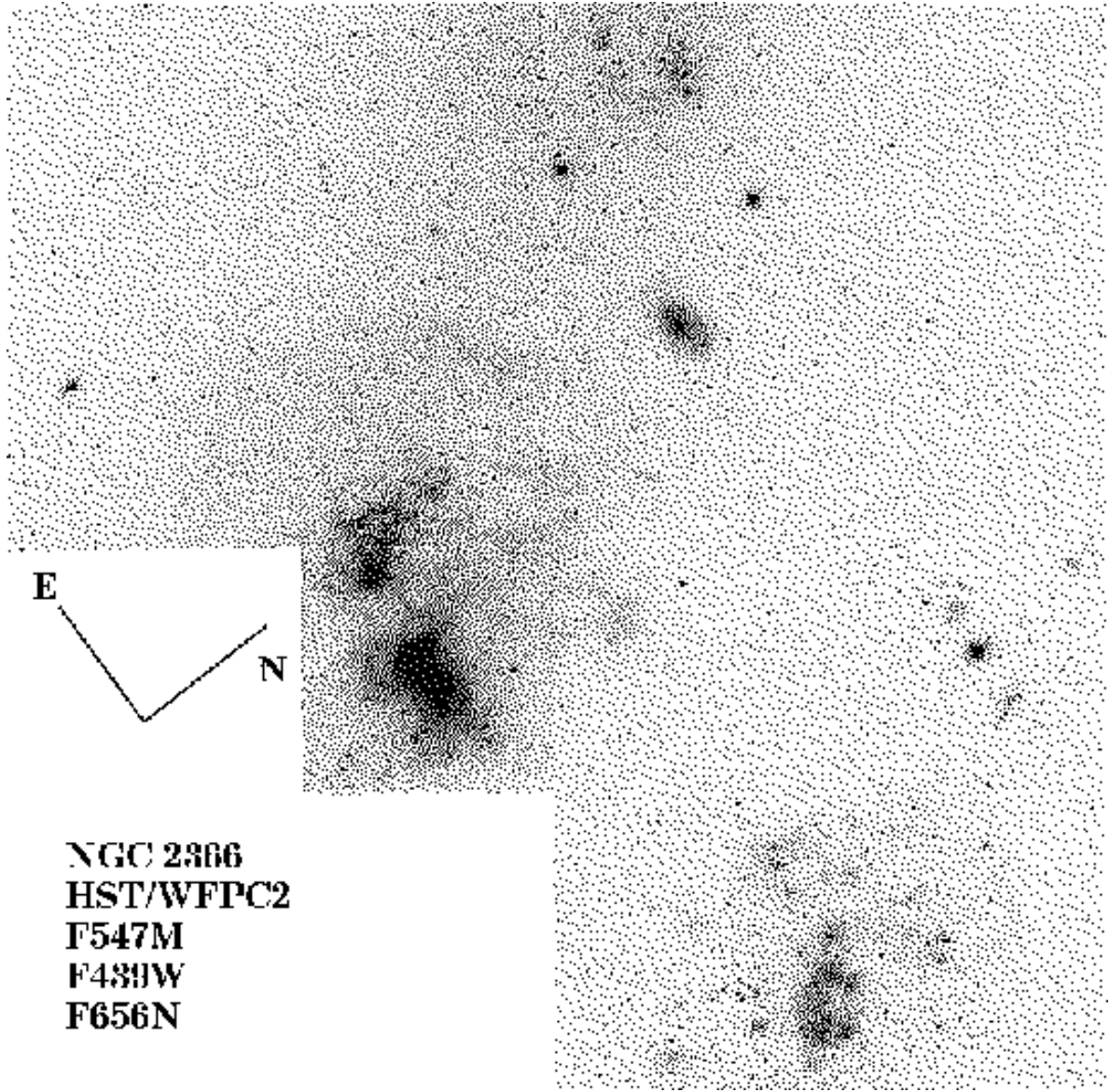}
\figcaption{Color-composite WFPC2 image of NGC 2366.
Full resolution color image available at http://astrosun.phy.ulaval.ca/astro/N2363.html }
\end{figure}

\begin{figure}
\figurenum{3}
\plotfiddle{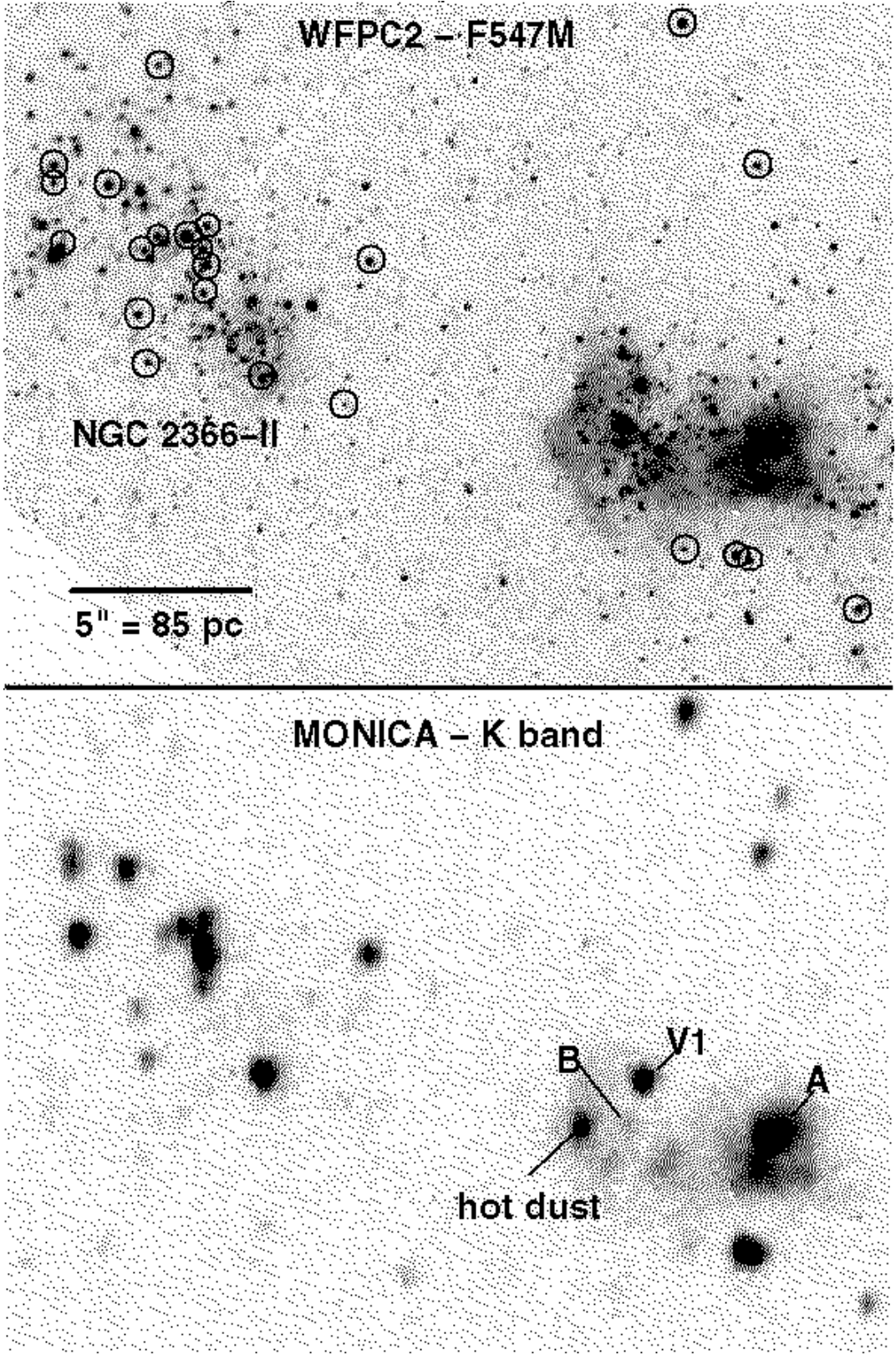}{210truemm}{0}{85}{85}{-245}{-25}
\figcaption{HST/WFPC2 and CFHT/Monica images
of NGC 2363(right) and NGC 2366-II (left). The red supergiant 
stars identified
in the color-magnitude diagram (in Figure 4, with $B - V \geq 1.0$ and
$M_V \leq -5.0$) are circled on the WFPC2 image; these are
prominent in the K-band image. Note that cluster A is very bright in K,
contrary to cluster B, as well as the possible presence of hot
dust (see text).
}
\end{figure}

\begin{figure}
\figurenum{4}
\plotfiddle{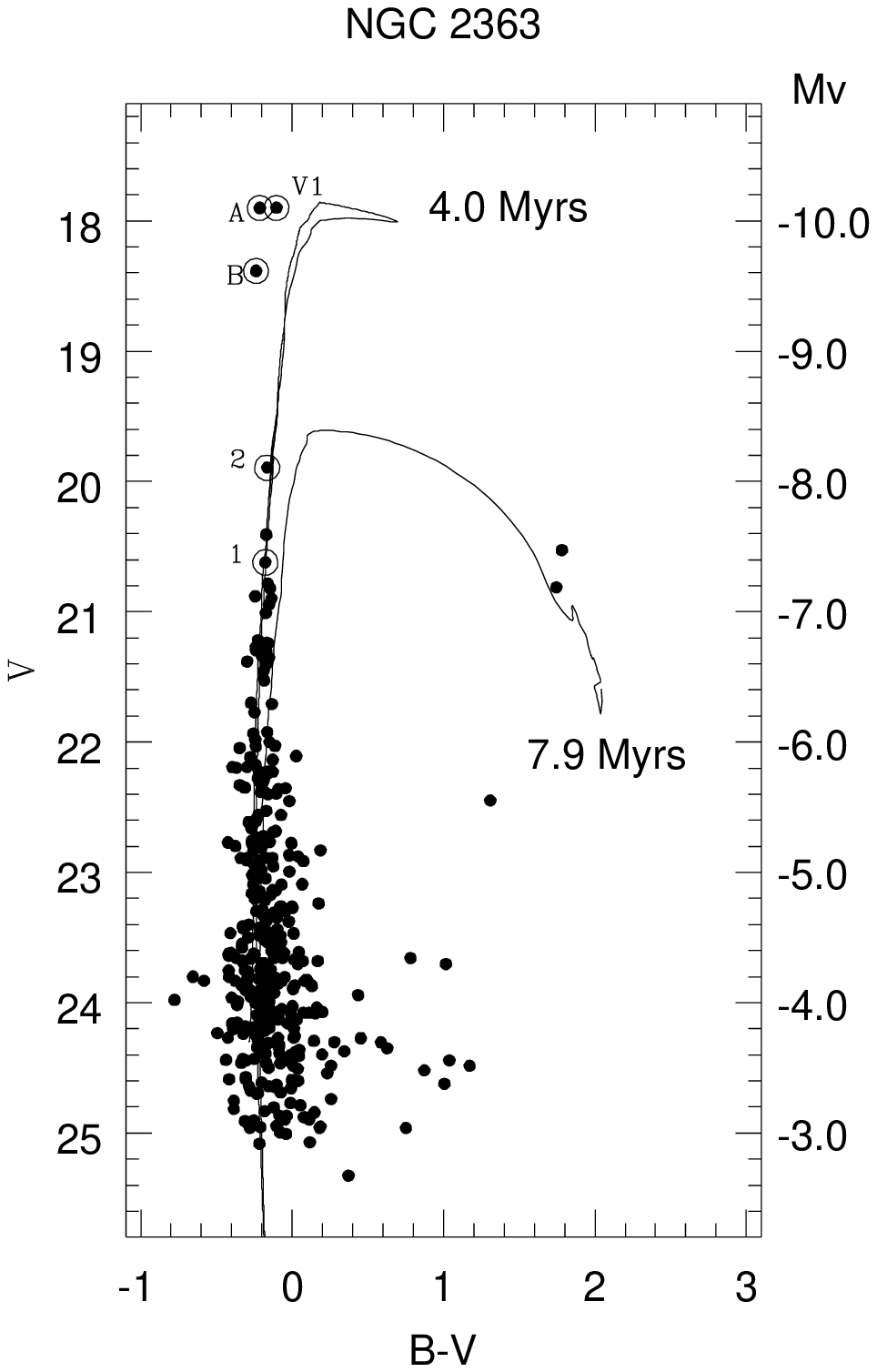}{100truemm}{0}{80}{80}{-290}{-203}
\plotfiddle{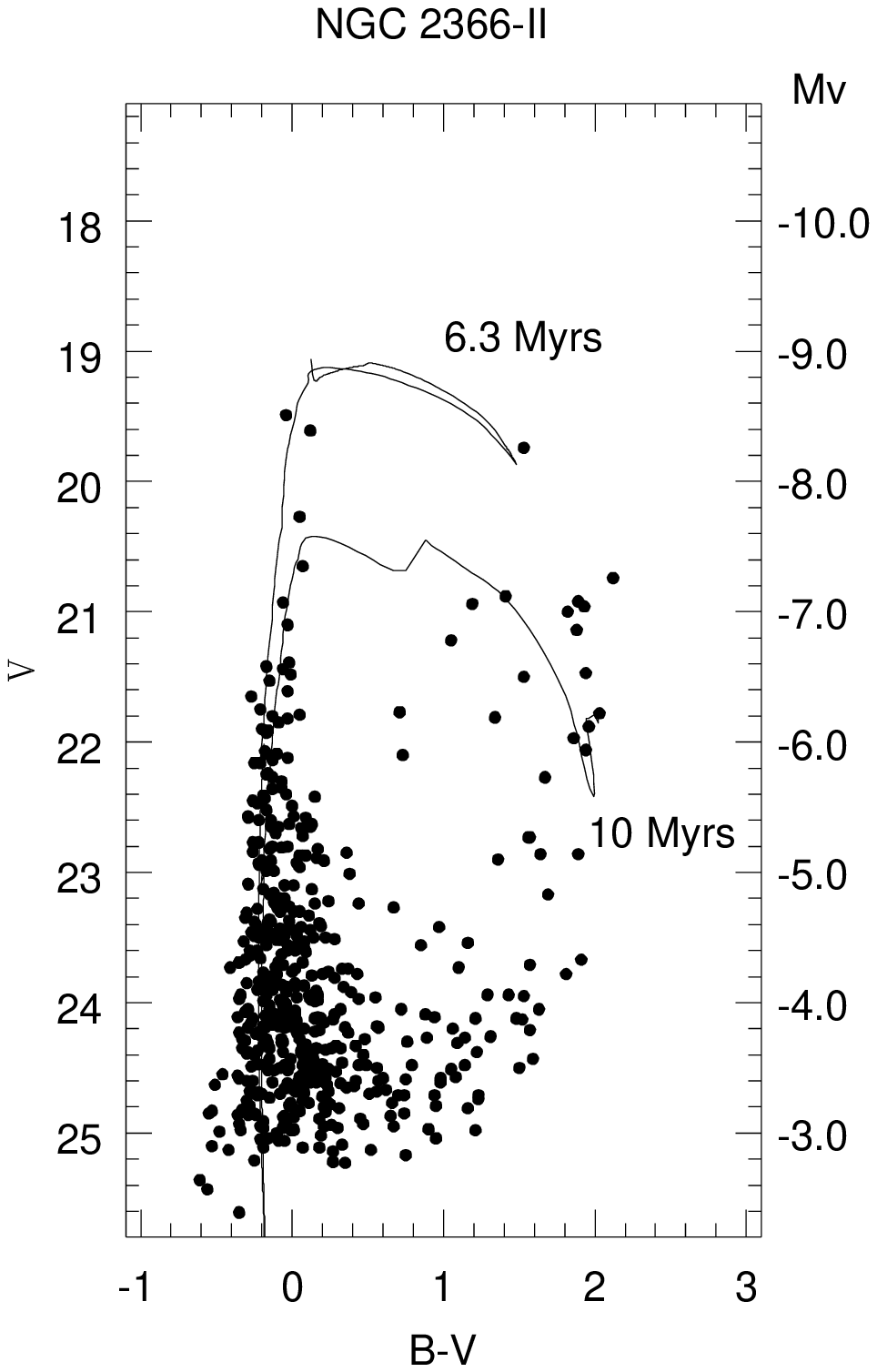}{30truemm}{0}{80}{80}{-60}{-100}
\figcaption{Color-Magnitude diagram of NGC 2363 and NGC 2366-II. The magnitudes
of clusters A and B in NGC 2363 are measured with an aperture of r = 0.43$''$
to match that of the FOS spectrograms.}
\end{figure}

\begin{figure}
\figurenum{5}
\plotone{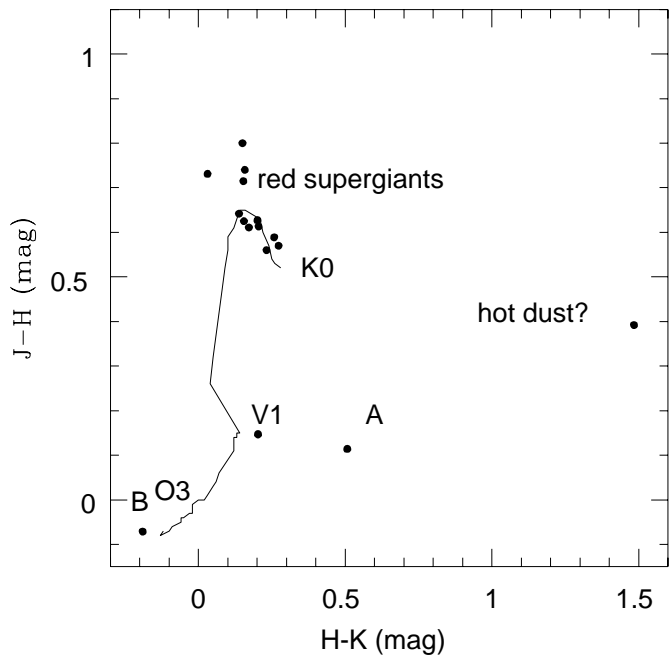}
\figcaption{Near-infrared two-color diagram of NGC 2363 and NGC 2366-II. 
The location of the supergiant sequence (from O3 to K0) is shown for
comparison. Knots A and B, as well as NGC 2363-V1 and the hot dust patch
are identified.
}
\end{figure}

\begin{figure}
\figurenum{6}
\plotfiddle{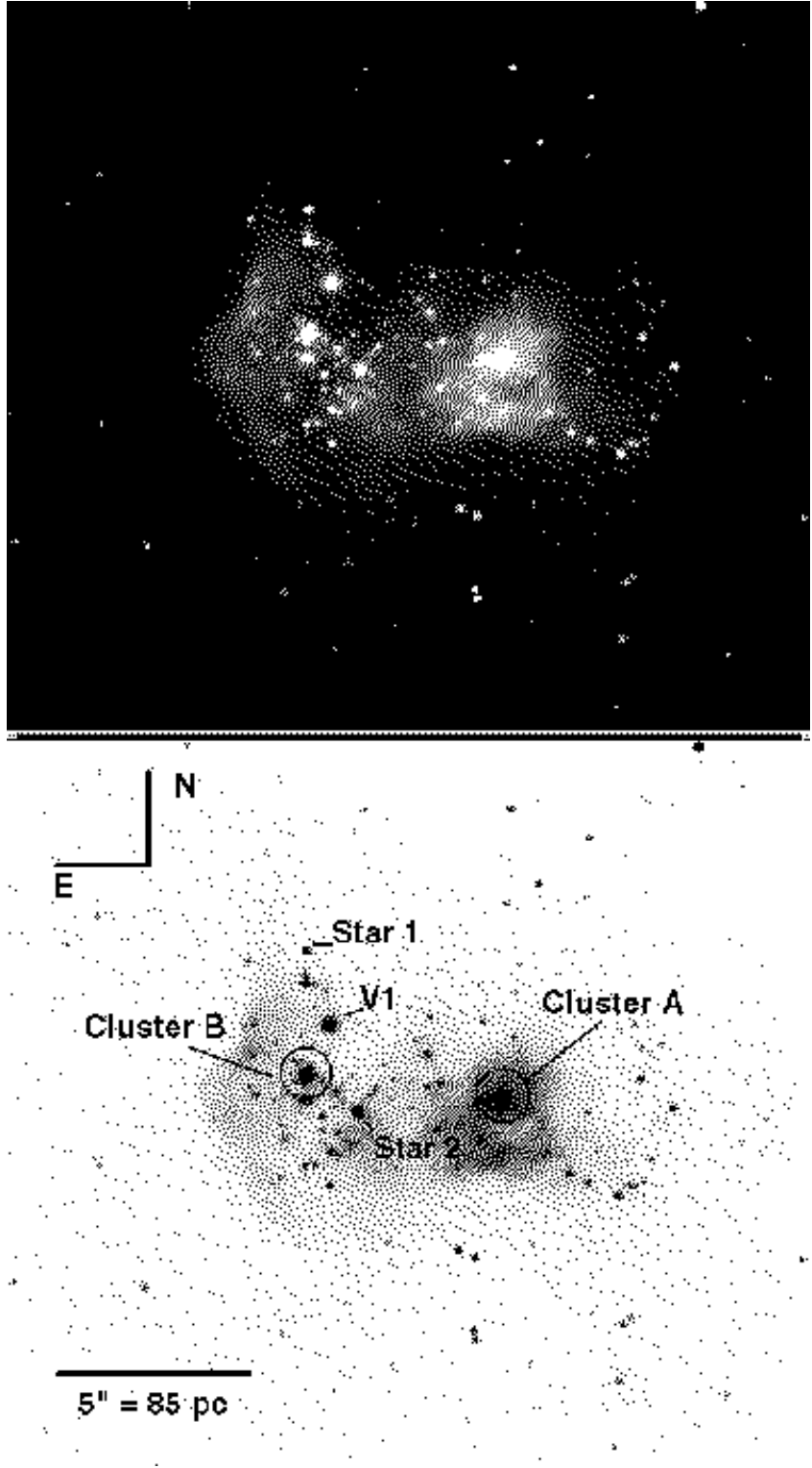}{200truemm}{0}{80}{80}{-245}{10}
\figcaption{[top] Color-composite WFPC2 image of NGC 2363 with the B (F439W) frame
in the blue beam, the average of the B and V (F547M) frames in the green beam
and the average of the V and $H\alpha$ (F656N) frames in the red beam.
The field of view is $21'' \times 19''$ ($350 \times 325$ pc). [bottom] Same image as above showing
the different objects described in the text; the two HST/FOS apertures are also
shown as circles.}
\end{figure}

\begin{figure}
\figurenum{7}
\plotone{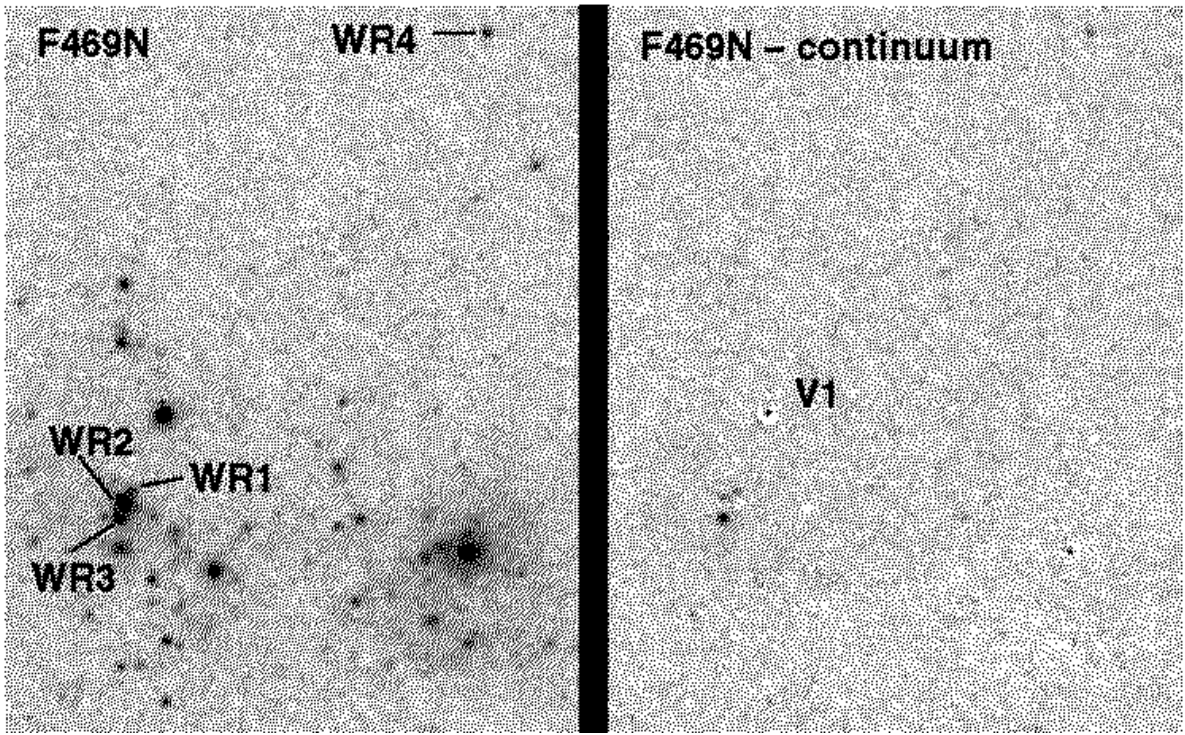}
\figcaption{F469N (He\,{\sc ii} $\lambda$4686) and net 4686 images of NGC 2363. 
WR candidates are identified. Note the positive and negative pixels at the
location of V1 and cluster A; these artefacts are the result of the 
imperfect image subtraction and show up because the two objects are by far
the brightest in the image, but they do not represent significant He\,{\sc ii}
enhancements. http://astrosun.phy.ulaval.ca/astro/N2363.html}
\end{figure}

\newpage

\begin{figure}
\figurenum{8}
\plotone{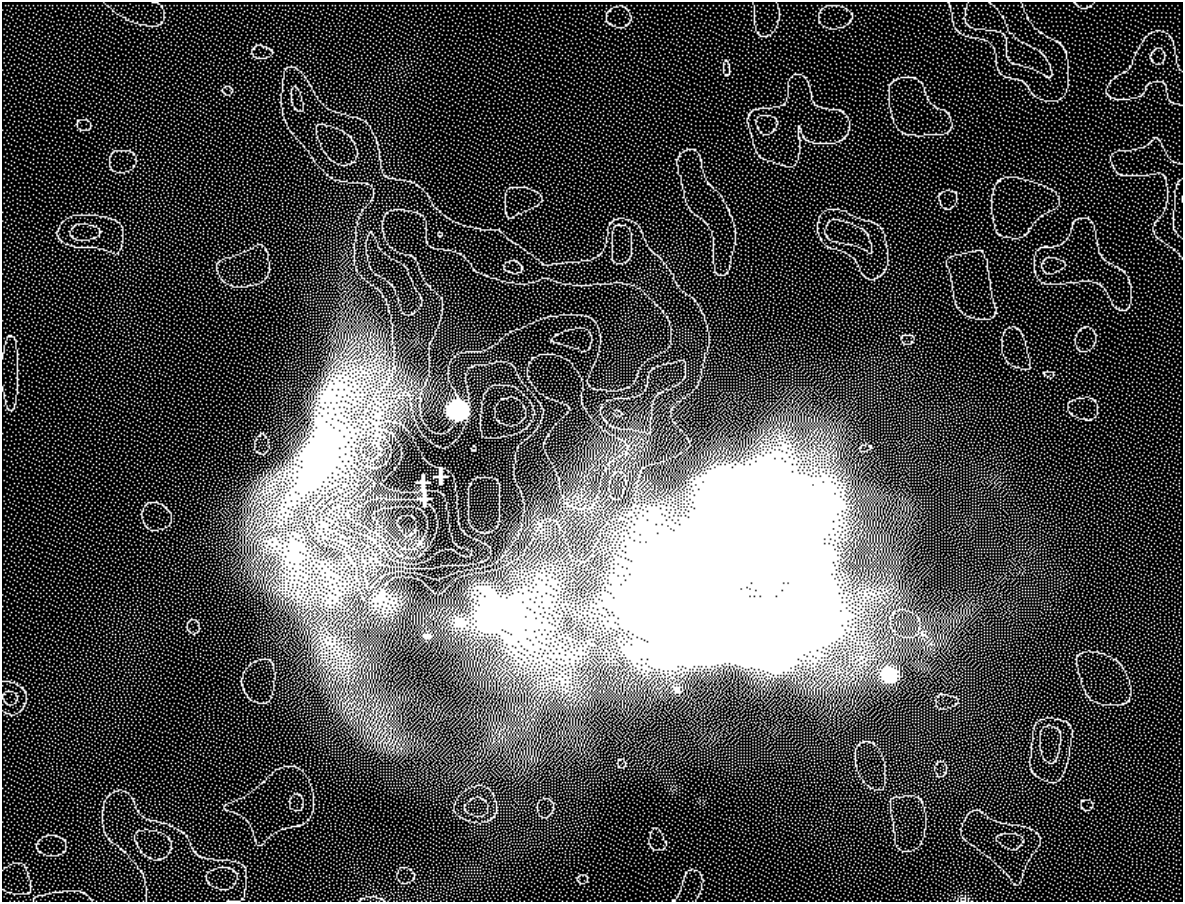}
\figcaption{H$\alpha$ image of NGC 2363, with contours of the He\,{\sc ii} 4686 diffuse excess. The point sources identified as Wolf-Rayet stars have been
removed from the net $\lambda$ 4686 image and marked here as plus signs.
The lowest contour corresponds to 0.5$\sigma$ (standard deviation from
the sky background), with increments of 0.5$\sigma$ per contour level up
to 0.35$\sigma$.
Note the spatial correspondance with the apparent H$\alpha$ cavity.}
\end{figure}

\begin{figure}
\figurenum{9}
\plotone{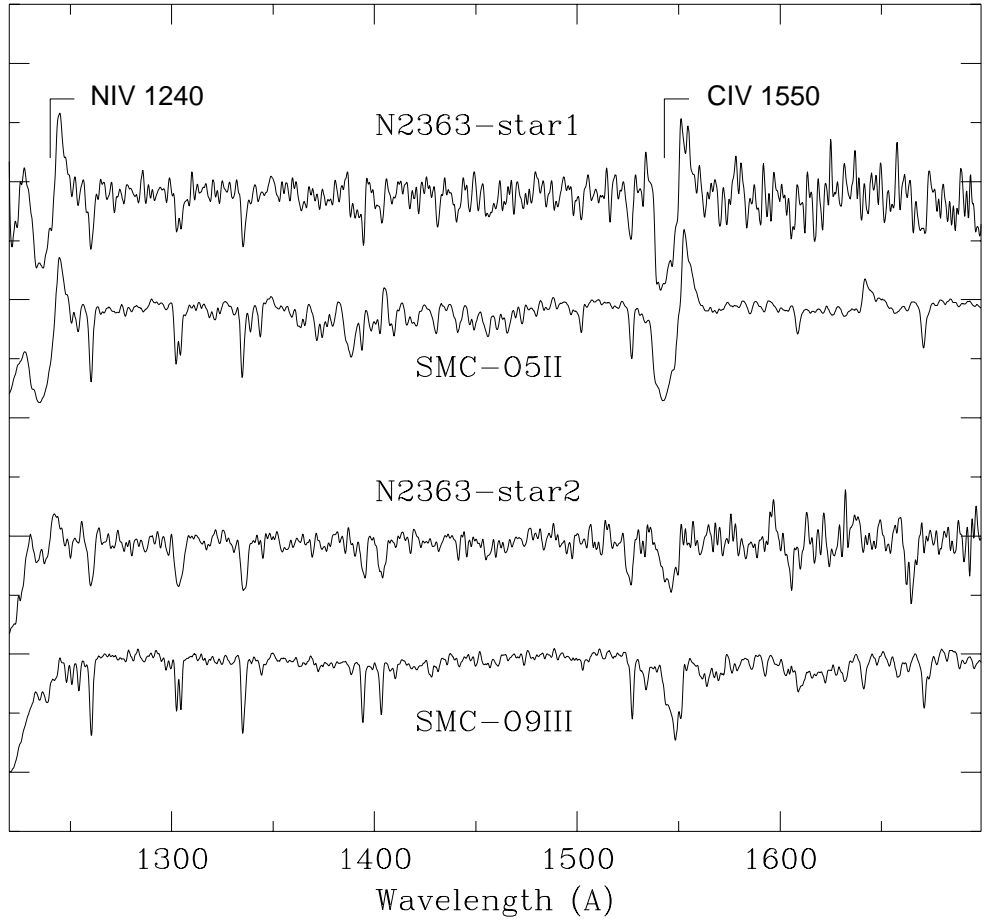}
\figcaption{STIS spectra of stars 1 and 2 (identified in Figure 6). For comparison, representative
spectra of stars with similar spectral characteristics in the SMC are
also shown.}
\end{figure}

\begin{figure}
\figurenum{10}
\plotone{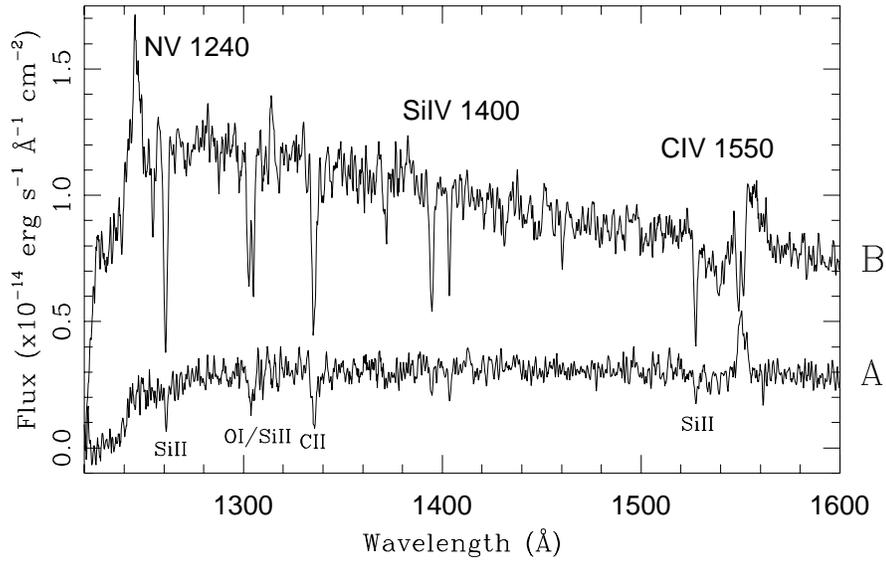}
\figcaption{Flux-calibrated HST/FOS spectra of knots A and B.}
\end{figure}

\begin{figure}
\figurenum{11}
\plotone{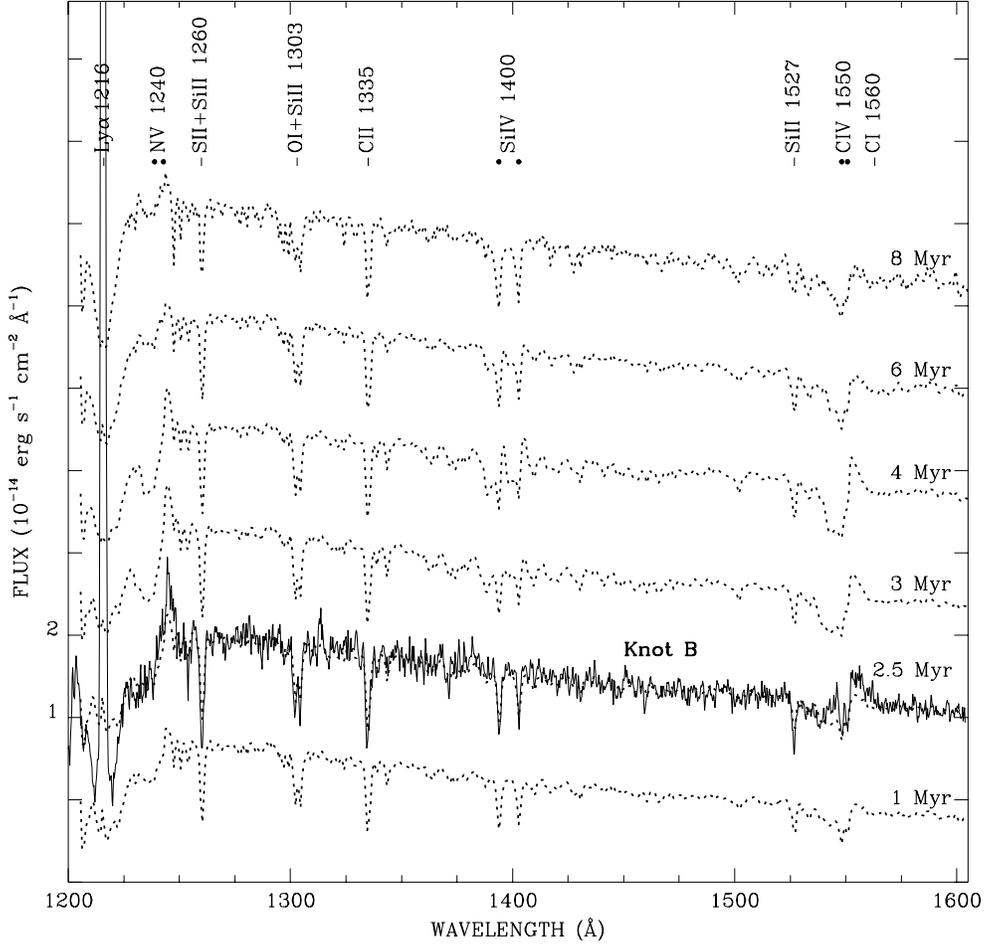}
\figcaption{Comparison of synthetic UV spectra for knot B. The spectra of
knot B (full line) is dereddened using $E(B-V)~=~0.058$ and an SMC reddening
law. Synthetic spectra (dotted lines) for various ages have been calculated
for an instantaneous burst using the stellar evolutionary tracks at a
metallicity $0.1~Z_\odot$, the SMC spectral library, and a Salpeter type
IMF ($\alpha~=~2.35$) with stars between 1 and 80~$M_\odot$. All the synthetic
spectra, except the one at 2.5~Myr, are shifted. 
The strongest interstellar and stellar lines are
identified at the top. A dot preceding the line identification indicates a
line formed mainly in hot star winds.
The synthetic spectrum at 2.5~Myr is the best model retained based
on the UV line synthesis.}
\end{figure}

\begin{figure}
\figurenum{12}
\plotone{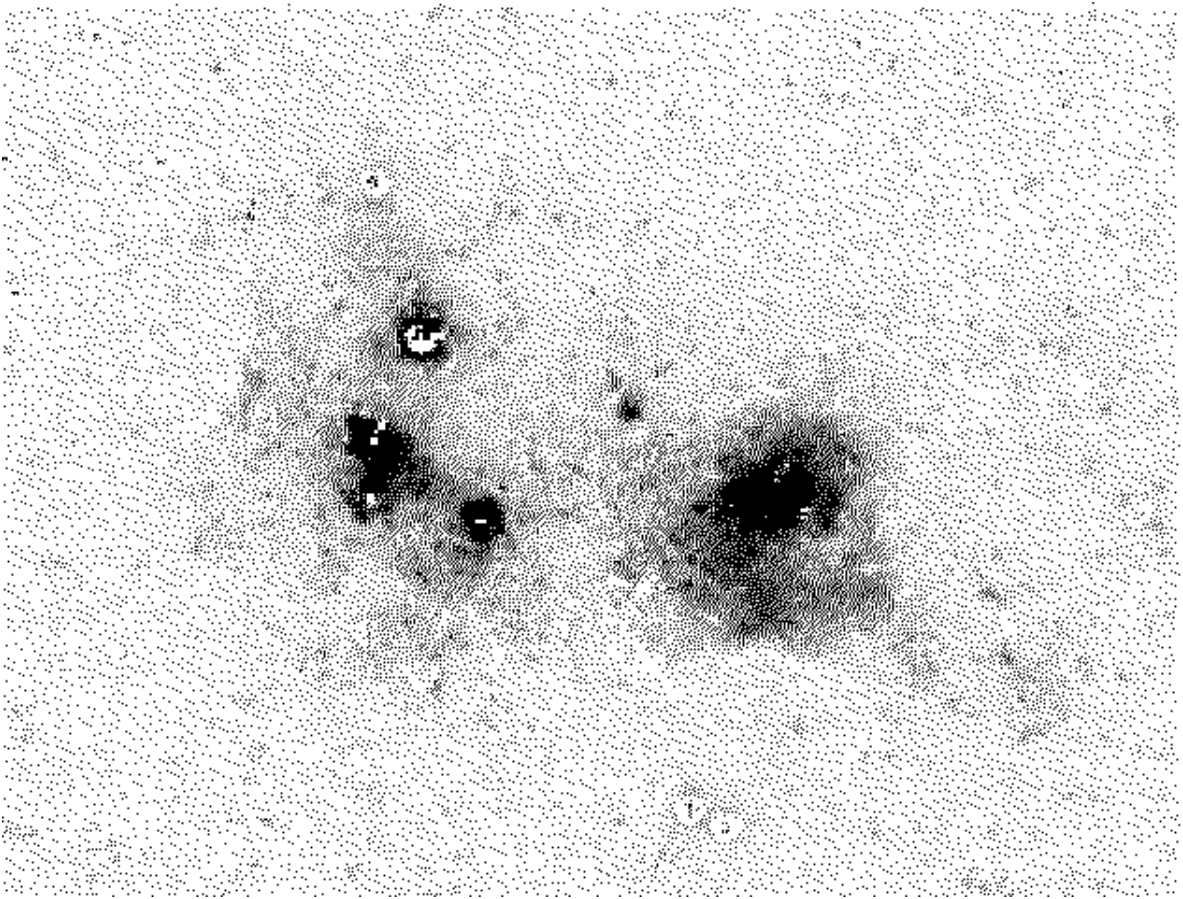}
\figcaption{Diffuse stellar light in the F547M WFPC2 filter scattered
by interstellar dust in NGC 2363. Display is in logarithmic scale. }
\end{figure}

\begin{figure}
\figurenum{13}
\plotone{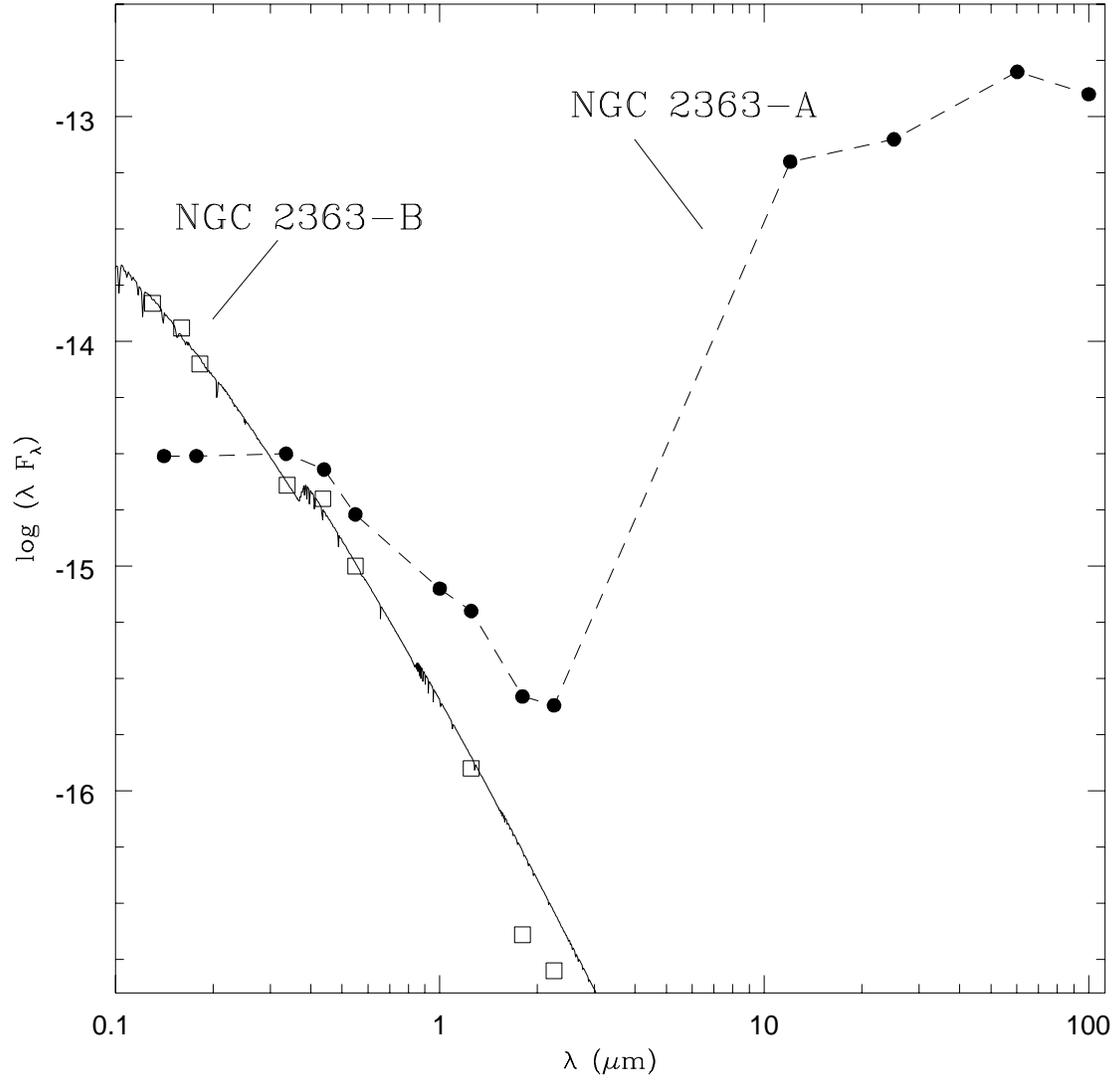}
\figcaption{Spectral energy distribution of NGC 2363-A and B. The solid line
joining the data points for NGC 2363-B is the energy distribution
of an instantaneous starburst
aged 3 Myrs. The dotted line joining the data points for NGC 2363-A is 
just intended to guide the eye; no attempt to fit a model distribution
was made.
}
\end{figure}

\begin{figure}
\figurenum{14}
\plotone{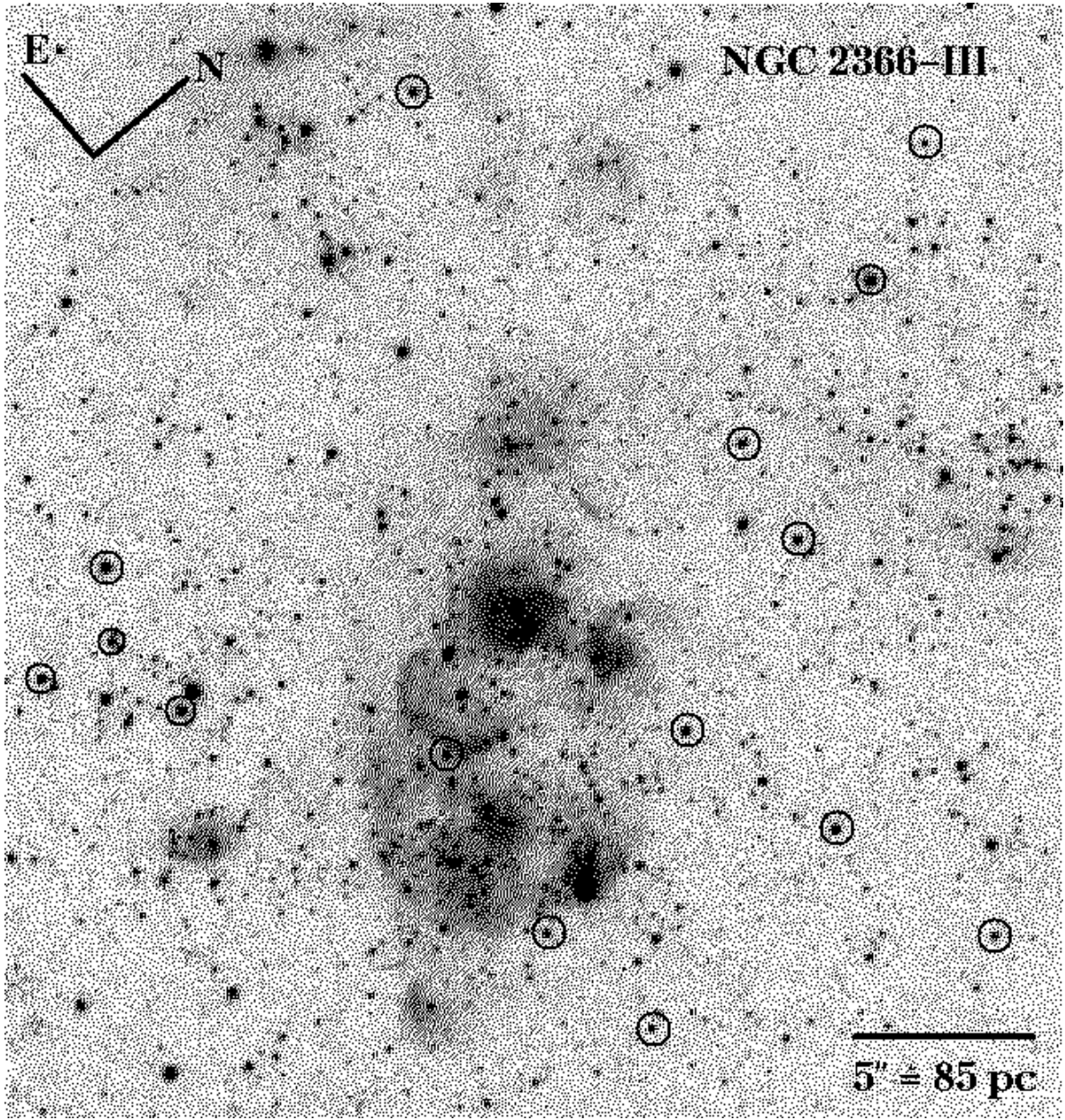}
\figcaption{WFPC2 image of the central region of
NGC 2366-III. Red supergiant stars are marked.}
\end{figure}

\begin{figure}
\figurenum{15}
\plotfiddle{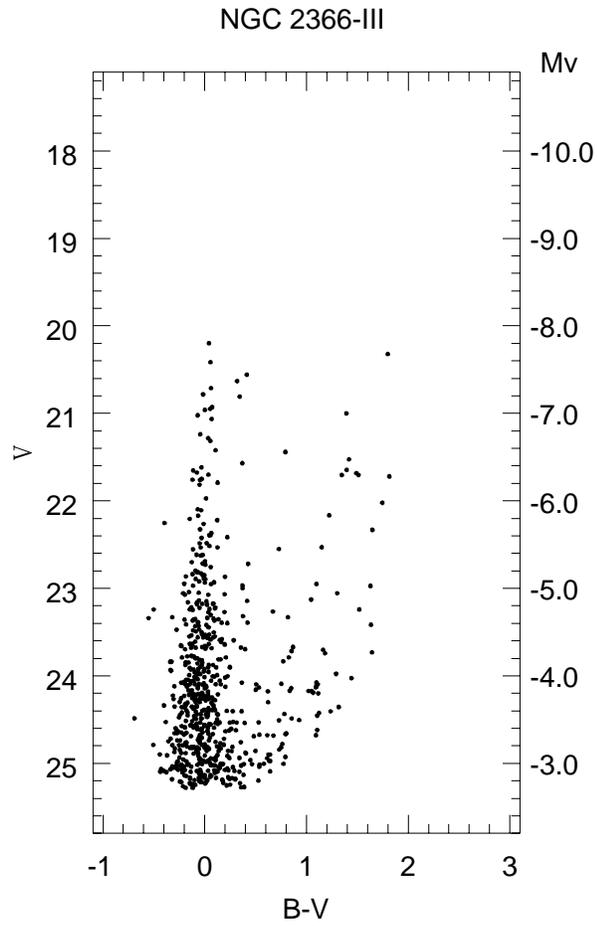}{100truemm}{0}{80}{80}{-250}{-153}
\figcaption{Color-magnitude diagram of NGC 2366-III.}
\end{figure}

\begin{figure}
\figurenum{16}
\plotone{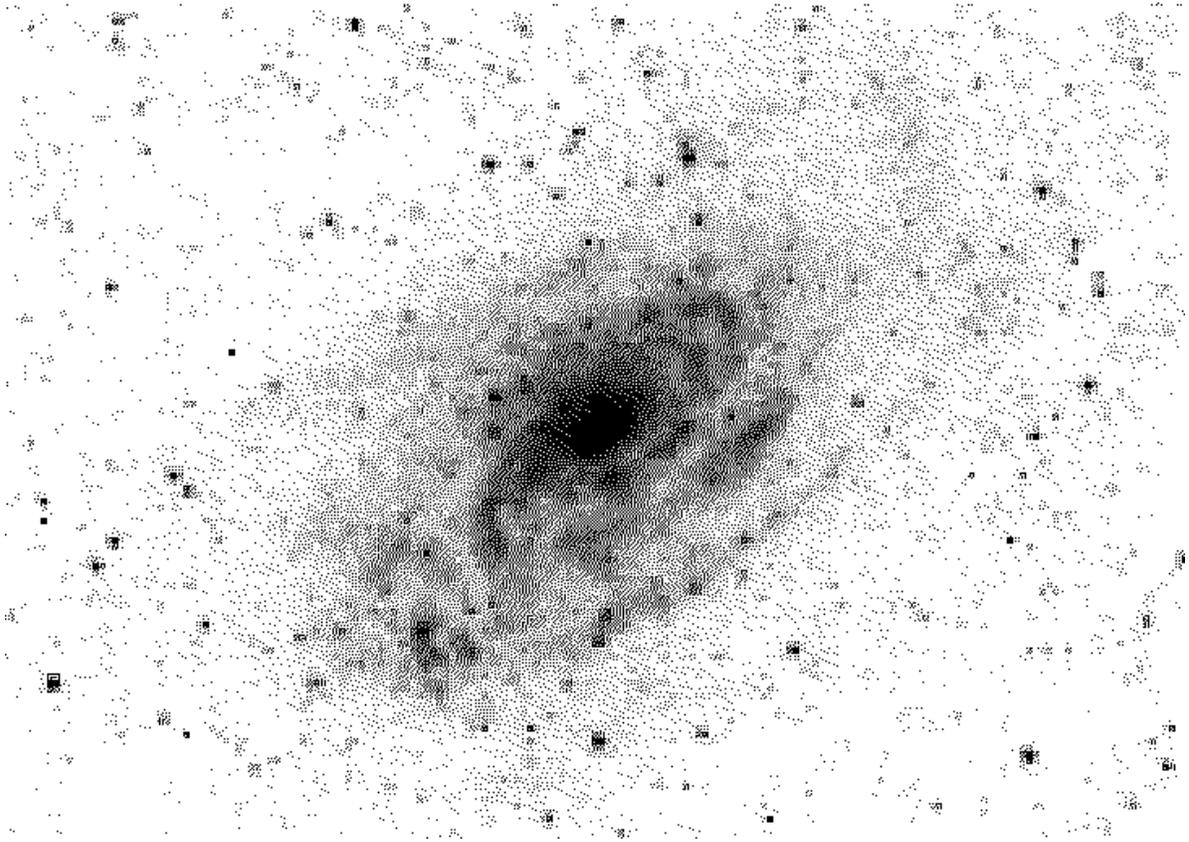}
\figcaption{Distant Sc spiral galaxy at $d \sim$ 500 Mpc seen through
the disk of NGC 2366. The scale of the image is 18 $\times$ 13 arcsec$^2$.}
\end{figure}

\end{document}